\documentclass[technologies,short note,accept,oneauthors,pdftex,10pt,a4paper]{mdpi}
\usepackage{amsmath}
\usepackage[utf8]{inputenc}
\usepackage[english]{babel}
 
\usepackage{amsthm}

\usepackage{soul,upgreek}

\usepackage{amsmath}
\usepackage{mathrsfs}

\usepackage{amsfonts}
\usepackage{amssymb}

 \usepackage{microtype}

\newcommand\be{\begin{equation}}
\newcommand\ee{\end{equation}}
\newcommand\bea{\begin{eqnarray}}
\newcommand\eea{\end{eqnarray}}
\newcommand\bseq{\begin{subequations}} 
\newcommand\eseq{\end{subequations}}
\newcommand\bcas{\begin{cases}}
\newcommand\ecas{\end{cases}}

\newcommand{\p}{\bfseries}
\newcommand{\m}{\mdseries}

\firstpage{1} 
\articlenumber{x}
\doinum{10.3390/------}
\pubvolume{5}
\pubyear{2017}
\copyrightyear{2017}
\history{Received: 7 July 2017; Accepted: 23 August 2017; Published: date}

\Title{\textls[-20]{Semiclassical Length Measure from a~Quantum-Gravity Wave Function}}

\Author{Orchidea Maria Lecian 

}

\AuthorNames{Orchidea Maria Lecian}

\address[1]{%
DIAEE---Department for Astronautics Engineering, Electrical and Energetics, Sapienza University of Rome, Via~Eudossiana, 18, 00184 Rome, Italy; lecian@icra.it}

\corres{Correspondence: lecian@icra.it; Tel.: +39-3392041294}

\abstract{
The definition of a length operator in quantum cosmology is usually influenced by a~quantum theory for gravity considered. The semiclassical limit at the Planck age must meet the requirements implied in present observations. The features of a semiclassical wave-functional state are investigated, for which the modern measure(ment)s is consistent. 
The results of a length measurement at present times are compared with the same measurement operation at cosmological times. By this measure, it is possible to discriminate, within the same Planck-length expansion, the corrections to a Minkowski flat space possibly due to classicalization of quantum phenomena at the Planck time and those due to possible quantum-gravitational manifestations of present times.  
This~analysis and the comparison with the previous literature can be framed as a test for the verification of the time at which anomalies at present related to the gravitational field, and, in particular, whether they are ascribed to the classicalization epoch. Indeed, it allows to discriminate not only within the possible quantum features of the quasi (Minkowski) flat spacetime, but also from (possibly Lorentz violating) phenomena detectable at high-energy astrophysical scales.
The results of two different (coordinate) length measures have been compared both at cosmological time and as a perturbation element on flat Minkowski spacetime. The differences for the components of the corresponding classical(ized) metric tensor have been analyzed at different orders of expansions. 
The results of the expectation values of a length operator in the universe at the Planck time must be comparable with the same length measurements at present times, as far as the metric tensor is concerned.  
The comparison of the results of (straight) length measures in two different directions, in particular, can encode the pertinent information about the parameters defining the semiclassical wavefunctional for (semiclassicalized) gravitational field.}


\begin{document}


\section{Introduction\label{section1}}
Quantum theories of gravity are usually implemented in cosmology for ages of the universe below the Planckian scale.
Any quantum effect for the gravitational field must be predicted as negligible after that age; correspondingly, the
classicalization of such effects has to be considered possible as far as such quantum-gravitational effects are involved
in the interpretation of current features of the~universe.

The understanding of the allowed description for the evolution of the metric tensor starting form the cosmological  singularity and undergoing the story of the universe is influenced by the quantum features potentially acquired by the gravitational interaction at the appropriate time, i.e., at the very early instants ofter the formation of the universe.

The classicalization of the quantum-gravity effects, both due to the gravitational interaction and to the presence of matter, is of importance, and the discrimination between the possible interaction patterns can be traced back within the experimental data at present times.

The modifications to gravity due to the presence of matter perturbations and of gravitational perturbations is such that the gravitational potential is not modified for perturbations allowing to reconstruct the present features of the universe.

The modalities of classicalization of perturbations can therefore be looked for in the anisotropy patterns individuated in both the features of the metric tensor and that of the distribution of celestial bodies, for which an alteration form a perfectly isotropic (spherically symmetric model also as far as the matter distribution is concerned) is required. In the present work attention will be paid to the features of the metric tensor, which can be ascribed to classicalization of perturbations, and which can define and shed light o the quantum-gravitational features of the gravitational interaction possible immediately after the Planck age.
The semiclassical regime of gravity can be described by a wavefunctional, which encodes the relevant information with respect to the solution to the (non-quantum) Einstein field equations. The metric tensor, which provides the needed tool for such a description, must encode the features of the growth of the universe, and the semiclassical wavefunctional must encode it at the given time. In particular, also the anisotropic expansion directions are supposed to qualify the wavefunction, corresponding to the appropriate quantum system, for which (also) a length operator is~defined.

The relic of the semiclassical age of the universe, during which also non-gravitational phenomena can be hypothesized to explain its present features, should be outlined in present experiment-like comparison, for which the contributions descending from different quantum (Minkowski) gravitational effects can be distinguished.   
The cosmological setting of a perturbed FRW model at the Planck scale is not only probable, but also of favored interpretation for the present universe, as far as experimental data are concerned, both for the empty sector (such as CMB fluctuations) and or the matter distribution (at galactic and extragalactic scales).

As far as the analysis of a classicalizing universe is concerned, the nature of the perturbations is unimportant, i.e., either of gravitational origin or by matter fields, as the classicalization phenomena can be described gravitationally after the Planck age, as a slightly perturbed FRW model. Within this phenomenon, several aspects can be included, such as the presence of matter after the cosmological singularity, and the role of matter in suppressing some cosmological models. Indeed, all the suppression mechanisms are based on the role of external fields in dominating the perturbation terms in the EFE's without modifying the Gravitational potential, i.e., without any term summing to the Ricci scalar at the considered orders in the pertinent (time) Taylor expansion.
 
The (classicalizing) quantum nature of the gravitational interaction at the Planck scale can be explored by the analysis of the behavior of quantum-mechanical operators; the length operator needs the GR notion of geodesics to be defined on quantum-gravitational states. The outcome of a~length measure is therefore related to the possibility of a (geodesics) path to be `transited', while the definition of quantum-gravitational states is related to the (geodesics, Planck length) segments on which quantum-gravitational states are defined.

The possibility to define a length measure at the beginning of the semiclassical epoch is therefore set by the change in the relevance of the quantum effects, which, at the beginning of the classicalization epoch, can be treated as perturbations. The definition of length measures at this stage of the evolution of the universe cannot therefore be affected by the specification of the perturbation mechanisms, which defines the classicalization, but only on the features of the metric tensor.

The perturbations of resulting from the classicalization of a quantum universe are relevant in defining the classical features of the present universe, which can be determined only at the corresponding (semiclassical) epoch, at which specific (possibly also strongly anisotropic) features are suppressed. The suppression mechanism can be established not only by the presence of matter in the Einstein equations, but also by its quantum features, which may also include extra gravitational terms in the Hamiltonian \cite{Lobo:2013vga}.

At the semiclassical epoch, the gravitational field is described by the solution to the Einstein field equations, while quantum-mechanical system by the proper Schrodinger equation. The semiclassical wavefunctional for the evolving universe must therefore encode all the features, by which the growth of the (an)isotropic volume of the universe is described, as well as all the matter contributions to the EFE's, i.e., including the presence of classicalizing matter fields.

\textls[-20]{From a different point of view, it has been pointed out that \cite{Ambjorn:2005db} a quantum universe appears as a~$1+1(^{n})$ dimensional universe, while its passage to the non-quantum phase renders its $4$ spacetime~dimensions.}

The result of measurement operations must be necessarily independent of the particular quantum theory of gravity adopted for the description of the quantum features of gravity at quantum times, i.e.,~before the Planck time, but must nevertheless encode the information about the features acquired but he gravitational field also because of the matter content of the universe. Such features, which (according to modern interpretation are aimed to) specify in the metric tensor the evolution of the three-volume of the universe, are not modified after the classicalization epoch, and must be still~detectable.

All the efforts aimed at qualifying the present anisotropy features of the universe notwithstanding, the result of a (straight) length measure(ment) operation still encode these features of the metric tensor. The semiclassical wavefunctional must contain this information (parameters), such that the results of a~length operation, which describe the classical effects of gravity, i.e., the metric tensor, must have the same unchanged particularization after the classicalized regime of gravity. This can be obtained by the ratio of the results of length measures performed for in (two) different (coordinate) directions.

\cite{Ambjorn:2005db}
The features of the General Relativity theory of gravity which can be requested to be outlined also within the quantum phase of the universe, such as the definition of a curvilinear abscissa, are outlined in Section \ref{section2}.
Length operators in quantum theories of gravity are reviewed in Section \ref{section3}, and~the limits for geodesics lengths are also discussed.

The techniques often used for the classicalization of the quantum phase of the universe are recalled in Section \ref{section4}, as well as the schemes needed for the definition of perturbed FRW universe.

Differently, the qualities of spacetime hinting quantum deviations from the Minkowski spacetime at present times are discussed in Section \ref{section5}, such as the possibilities to prepare (interferometer) experiments on newly-emerged (quasi)-Minkowski spacetimes.

\textls[-15]{Observational tests for quantum features of the spacetime arising at most at the semiclassicalization epoch and still detectable in gravity experiments are discussed in Section \ref{section6}.} In particular, the parity properties of the expectation value for a length measure are recognized in the dispersion relations.

A comparison between the possibility to reconstruct the components of the metric tensor from observations at present times at astrophysical scales is presented in the straightforward Hamiltonian limit.

The results are compared in Section \ref{section8}, where the possible modifications to the components of the metric tensor and their experimental characterization are discussed. In particular, the expansion for the components of the metric tensor for an almost-flat Minkowsky spacetime at present times is compared with those for a perturbed FRW spacetime at the Planck epoch.  

Brief concluding remarks follow in Section \ref{section9}.
    
\section{(Quantum) Geometrical Descriptions\label{section2}} \vspace{-7pt}

\subsection{The Heat Kernel for the Gravitational Field} 

The properties of the heat kernel for gravitational phenomena have been compared in
\cite{carlip2009}
 as far as the implication
for phenomena relevant at the Planck scale are concerned, being $l_P$ the Planck length and $t_P$ the Planck time..

The heat kernel $k$, $k = k(x, x_0; s)$ for the diffusion process of a particle from $x$ to $x_0$ (time \emph{t}) within the gravitational field and back (time $s$)  
obeys the differential equation
$
\left(\tfrac{\partial}{\partial s}-\Delta x\right)k=0
$
and is dimension-dependent only as
\begin{equation}
k(x, x'; s)\sim(4\pi s)^{-d_S/2},
\end{equation}
being $d_S$ the number of dimensions (and that of the spherical region considered).

In the Hamidew expansion,
\begin{equation}
k\sim\frac{1}{4\pi s^2}\left(a_0+a_1s+a_2s^2+\dots\right)
\end{equation}
the $a_1$ term, proportional to the scalar curvature, vanishes only for fact vacuum solutions, while, in~presence of
matter, the contributions to $a_1$ are diverging.

This is reflected also in the
\cite{futa84} Futamase-Berkin expansion,
\begin{equation}
k\sim\frac{1}{4\pi s^2}\left(1+\frac{a}{t^2}s+\dots\right)
\end{equation}
for fixed $t$, there always fists s small enough for $1$ to dominate, while, for fixed $s$, there can always fist a~time $t$ for the
scalar-curvature term to dominate.

This analysis notwithstanding, a notion of `geodesic radius' in $1+1$ quantum gravity has been discussed in \cite{Ambjorn:1998pz}, and its scaling (evolution) in time has been controlled numerically.
	\subsection{Semiclassical (Pre-)Geometry}
The models of quantum information can also be employed in the description of the evolution of the space degrees of freedom of an expanding universe, whose classicalization via decoherence is depicted in \cite{Zizzi:1999sx}, where the thermal history of the semiclassical universe is encoded in the appropriate Fock space as far as the space degrees of freedom o a deSitter universe is concerned. Within this framework, the vector degrees of freedom of the gravitational field and its scalar ones are analyzed in~\cite{Roh:2000na} as far as the definition of the fundamental (Planck mass $M_P$) scale is concerned.

Within this viewpoint, the statistical analysis of General Relativity (including Lorentz symmetry) leads to the evolution of states towards semiclassical states, whose semiclassical, low-energy limit corresponds to gravity condensate states, whose \cite{Major:2001ka} lowest energy level corresponds to a description with $j=1/2$.

Without referring to (the classicalization of) `pregeometry', the classicalization epoch of a quantum geometry in presence of a scalar field has been investigated, which is in line with the definition of quasi-isotropization, classicalization of perturbations for scalar fields a the same universe age and cosmological setting.

Before the Planck age, the hypothesis of quantum features for the space geometry needs the modification of the phase space for General Relativity, whose cosmological limit is a modification of the metric tensor according to the so induced `parameter of non-classicality'. The parameter of non-classicality $k$ modifies the (FRW) scale factor by a momentum label, such that $\partial\bar{a}^2/\partial k^2=0$ in the classical limit \cite{momlab}. The departure from classicality is therefore measured by (a suitable function of the momentum label) $k$ at small $k$. The emergence of a cosmological definition for the metric tensor is motivated by the cosmology interpretation of the scalar field \cite{momlab2}. The presence of   `test quantum fields' therefore allows for the substitution of operators to their mean square value, and allows for the definition of semiclassical trajectories at the Planck age on a (quantum/perturbed) isotropic geometry corresponding to an FRW metric.

\section{Length Operators\label{section3}}
Starting from the Planck age, the classicalization of perturbations
\cite{guth81,lind83},   $\mid \ \ \rangle_{\bar{\delta g}}$  must be consistent with the age and length scale at which QG effects are due. This is achieved by selecting the relevant contributions from the Hamiltonian. The corresponding quantum states encode information about the solution to the Einstein field equations and about the initial conditions and
must be consistent with the age and length scale at which QG effects are due. This is achieved by selecting the relevant
contributions from the Hamiltonian. The corresponding quantum states encode information about the solution to the
Einstein field equations and about the initial conditions.
The~same information items must be encoded not only in expectation values of operators, but also in its expression
when evaluated on quantum (matter) settings, such as a length operator for non-geodesic~curves.

The same information items must be encoded not only in expectation values of operators, but~also in its expression when evaluated on quantum (matter) settings, such as a length operator for non-geodesic curves
\begin{equation}\label{quasim}
L\sim\int_0^1\left(q_{ab}(c(t))\dot{c}^a(t)\dot{c}^b(t)\right)^{1/2}d^4x\\
\end{equation}

The results of measure operations must be independent of the definition of operators but consistent with the matter
content of the theory.

Given the symmetries of the solution to the Einstein field equations, on which metric tensor the connections are
defined, be $\psi_{\Gamma, c}$ a (cylindrical) function of the given group symmetry for the~connections.

The most direct operation possible within this model is the calculation of the expectation value of the elementary
length operator $L$ for a wedge $\omega$ on a quantum state  $\psi_{\Gamma, c}$. Such length scales as
\begin{equation}\label{elemlen}
\left(\psi_{\Gamma, c}, \hat{L}\psi_{\Gamma, c}\right)=\sum_{k, h}c^*_k(j_0)(L)^hc_k(j_0)\sim\sqrt{j_0}
\end{equation}
where the first-order expansion is obtained from numerical study
\cite{bianchi1997}.

Furthermore, given two curves, $\gamma$, $\gamma_1$, the two operators measuring their lengths do not commute if the two curves
intersect or almost intersect
$\left[\hat{L}_\gamma, \hat{L}_{\gamma^1}\right]\neq0$, $t_\gamma\cap t_{\gamma^1}\neq0$, 
while they commute if the curves do not intersect 
$t_\gamma\cap t_{\gamma^1}=0$.

The measure of geodesics does not differ form the implementation of this very same paradigm, for~the information on the metric tensor naturally contained in the connections, just as in GR. The~interest in this approach is the analysis of the (possible) quantum features of the spacetime.

For the measure of non-intersecting curves, above the Planck length, the state  $\Psi_c$ can be compared to the `heuristic'
state-
described in
\cite{pulli99}
, for which a discrete approximation at flat geometry leads to the expectation value for the metric tensor (upgraded operator)
\begin{equation}\label{pulli}
\langle q_{ab} \rangle=\delta_{ab}+\mathcal{O}\left(\frac{l_{Pl}}{\Lambda}\right)
\end{equation}
with $\Lambda>l_{Pl}$ a characteristic length scale. For cosmological investigation, such a characteristic length scale is naturally identified as the Planck length.

From a mathematical point of view, the passage from a discrete $1+1$ Minkowski lattice to (more complicated) spacetime(s) is described in \cite{Bombelli:2006nm,Saravani:2014gza} for higher dimensions; the different applications for singularities in General Relativity are analyzed in \cite{Hedrich:2009pb}.

Given a graph, with boundary conditions and definitions of assigned (positive) lengths, it~can be turned in a `quantum' graph by considering all the possible lengths attributed to a `side' under the specified vertex condition \cite{quantumgraphs}.

The localization of a particle within a tetrahedron can be discussed both in Special Relativity and in General Relativity, where the assumption on its localization, i.e., wrt the eigenfunctions of the differential operators, changes as the structure of the phase space is different for different differential operators \cite{loc}. The probabilities for adding vertices to a quantum graph are evaluated numerically in~\cite{Krugly:2011np}
without taking into account the natural scale of the Planck time.

The solutions to the Einstein field equations are to be considered valid after the Plank age.
At~the Planck age, the General-Relativity configuration on which quantum effects can be considered as classicalizing
perturbations can be chosen according to the features which are still observed at present~times. In particular, any such solution has to be requested to be compatible with the definition of lengths (almost) everywhere.

\subsection{Semiclassical States and Geometrical Analysis of the Gravity Quantum Background}
The definition of a length operator for a quantum theory of gravity is not unique. 

The definition of (the wavefunctions for) states for polyhedra (from which the definition of a state on a wedge can follow) is regulated by the request that the coefficients in the wavefunction efficiently selects the (geometrical) features of the wedge \cite{Rovelli:2006fw}.

The selection methods of a given wedge allow compare how the construction of a wavefunction on a wedge is defined by the number of dimensions (of the polyhedron to which the wedge is defined)~\cite{Livine:2006ab}. The (semiclassical) quantum-gravity effects emerge as depending on the vertex amplitude \cite{Rovelli:2005yj} rather on other aspects of polyhedra, for which the definition of elementary geometrical operators is allowed.

The area \cite{area} operator has been proposed as well as for the exploration of a -preferably, also, large- (two-dimensional) subregion of the quantum spacetime \cite{Hamma:2015xla}, with the aim to recover a physical description of the spacetime subregion within its semiclassical limit. Its expectation value has been demonstrated to be consistent also for a single-link state \cite{single} wavefunctions. Within the semiclassical limit, for the single-link semiclassical state, there exists an entire family of deformations of such state, which are still the ground state of the Hamiltonian.

The interest in the study of a three-dimensional subregion of the quantum spacetime at the semiclassical limit has also been stated in \cite{schlie1}, where a small volume of the semiclassical spacetime has been proven to be investigated by one volume operator, while a broader volume has been depicted by a set of such volume operators. In \cite{schlie2}, the large-tetrahedron limit for the volume operator has been analyzed as consistent for the definition (by its derivative) of a momentum operator, whose quantum numbers also define the expectation value of the volume operator.

The definition of polyhedra \cite{Bendjoudi:2016gom} has been shown to be possible within a geometrical analysis, for which volumes and areas are consistent, as well as the definition of normal lengths; the definition of volumes and area operators is also proposed in \cite{bian1009}. All these objects and operators are invariant under the action of the diffeomorphism operator, as assured by construction.

\subsection{Length (Pre-)Operators}

	As one fundamental object in General Relativity, connected with the possibility of a definition for a (n at least semiclassical) length operator can be recognized as (the square root of) the determinant of the metric tensor. The properties of this eigenvalues have been compared in \cite{Loll:1995wt} as well as its lattice discretization(s) and continuum (limit(s)) versions.
	
	The continuum limit for a length operator has been discussed in \cite{Thiemann:1996at} as the pertinently corresponding part descending from the determinant of the metric tensor. At a semiclassical level, a~hierarchy for the relevance of intersections in the evaluation o lengths of two intersecting curves had been fixed: the length(s) measure can be a challenging test for the classicalizing stage of quantum theories of gravity.
Indeed, it is different from the measurements of lengths in modern-times lab experiments performed on quasi-flat
spacetimes, as it has to be implemented on strong-gravity regimes characterized by \cite{llft} the $t^{-2}$ time-dependence in the
expansion of the Ricci scalar.

The well-definedness of the scalar product in the Minkowski theorem within all these specifications has been checked in \cite{Loll:1996nk}.

Following the determinations of LGQ, the length operator can admit also other definitions, for its expression prepared for a (wave)function cylindrical with respect to a graph, i.e., wrt a segmented path, on which edges are specified. As explored in \cite{wong,Ma:2010fy}, (at least) two different strategies can be followed, for such a purpose. As a derivation of the volume operator, the action of a length operator on such a function is shown to depend on the function and to carry information about the (symmetry groups which define the metric tensor, and, therefore, the variables characteristic of this theory, which account for the) background geometry. The definition of the same operator from a quantization of (\ref{quasim}) contains the same information about the (symmetries of the) metric tensor.

The problem of a quantum horizon for a metric is discussed in \cite{ansari}, as well as the related area fluctuations; this issue, however, is not applied to the isotropic growth of the space volume for a metric tensor at the Planckian age in cosmology. The so-called 'degenerate' spectrum, that corresponds to lower-dimensional geometrical objects (such as edges and vertices) lying on the surface delimiting a volume has been computed according to up to a regularization in \cite{frittrov}. Including (a parameter proportional to the Planck length controlling semiclassical) fluctuations of the metric shows that the area of a surface delimiting a volume is not an eigenstate of the area operator any more \cite{ansari2}.

\section{Classicalization Techniques\label{section4}}
The definition of a length operator is therefore consistent within the characterization of gravity at the quantum scale. The number of dimensions naturally naturally characterizes the features of the gravitational interaction. The description of the quantum geometry is encoded within the definition of polyhedra, on which physical observables are defined    \cite{1999}.

The definition of a length operator at the semiclassical level descends therefore from the analysis of the geometrical description of the quantum geometry; the pertinent wavefunctions are shaped according to the (mathematical) properties of the background geometries. More in particular, the length operator must not be implied by choice of the classical geometry on which quantum-gravitational geometries are constructed. The most probable semiclassical transition can be found by minimizing the generating potential of the scalar field responsible for the transition with respect to volume \cite{calzetta2}. The general-relativity limit has therefore to be estimated according to the physical setting (i.e.,~time of the story of the Universe) at which the limit is considered.

From all the quantum theories of gravity which have been studied also in their cosmological version, the investigations
in `loopy' theories of gravity have long been focalized on length measurements, as the pertinent quantum states
encode the symmetries of the metric tensor, while the `generalized segments' whose lengths has to be measured
must depend on the definition of a~geodesics path as far as an 'observationally-motivated' approach is concerned.
More in particular, the quantum states (or wavefunctions) present in the quantum description of gravity contain the
information pertinent not (only) to the solution to the Einstein field equations, but to its symmetries. For this, the~definition
of  `generalized segments' is needed, as it is possible, within quantum theories of r gravity, to describe the measure of
lengths expressed by geometrical entities which are not concerned with the notion of geodesics, whatever the solution
to the Einstein field equations.

\subsection{Geometry before the Planck Scale-Comparison with Pregeometry}
The 'pregeometrical' description of spacetime, i.e., a description where no complete geometrical features for spacetime are available, because of the quantum scale, is motivated in \cite{wheeler}.

The concept of 'pregeometry' has been pointed out to consist of entities, which are not of a~geometrical character, and to lead to the classical notion of spacetime geometry by the continuum limit of their relations \cite{lorente}.

The pregeometric structure of spacetime, according to its present-time flat limit, has been described in \cite{Stuckey:2000ps}, where events are interpreted as responsible for the appearance of the geometrical features of spacetime.

A pregeometric theory for the big-bang has been formulated in \cite{friedmann} ,based on events represented by random graphs, whose creation and annihilation generates the geometrical features of spacetime.

The pregeometry of the (hot) big bang has been described in \cite{terazawa0}, where all the possibilities for the gravitational interactions have been considered. The quantum fluctuations that characterize the transition from a pregeometrical gravitational field to a geometrical gravitational field have been reviewed in \cite{terazawa}, where an auxiliary (scalar) field is also introduced, accordingly.

A pregeometric metric quantum space has been proposed in \cite{alvarez}, where the quantum variable considered corresponds to the quantum phenomenon of the distance between two points; the resulting quantum space for the quantum states has features is common with a manifold. The quantum error (indeterminacy) corresponding to these measurements has been calculated in \cite{ngdam}.


\subsection{Linearization Techniques for Planck-Age Anisotropies}
The quasi-isotropization mechanisms acting on an isotropic cosmology are model independent~\cite{futa84} (i.e., independent
of the particular scalar-field gravity coupling of the resulting toy-model \cite{bel73}) for the cold phase of any (also strongly)
anisotropic Early Universe.

The linearization of homogeneous quasi-isotropic cosmological
\cite{pontz2009}
 models is consistent with perturbations of an
FRW model, for which the contributions to the potential are (factly) second-order for the field equations, such that
the potential term R stays unchanged also for a change (not always a rescaling) of the (group) structure constants
for the symmetry group characterizing the space part of the metric tensor (solution to the Einstein field equations).
After this analysis, one learns that, at~a~given time instant at (or immediately after) Planck age, the (realistic,
quasi-isotropic) cosmological picture obtained from Bianchi cosmologies is also depicted by perturbations of an FRW
universe also as far as the symmetries of the metric tensor are concerned, i.e., without changes nor for the curvature
scalar or for its time expansion. As far as cosmological investigation is concerned, the corresponding cosmological
evolution(s) is obviously different: following the (proper)-time evolution of the metric tensor, the determination of the time (age of the universe) at which possibly strong anisotropies are begun to be modified is needed.

This simplifying assumption partially reconciles the difficulties of the thermalization process
\cite{dor71}.

The description of the appearance of spikes \cite{lars2004}
within the oscillatory regime can be consistent, at~each time,
with a change (not always a rescaling) of the structure constants for the symmetry group characterizing the space
part of the metric tensor.
The description is not only consistent with the linearization technique, but also suited for the investigation of
the behavior of QG length operators.
As a different result, the expansion of the result of a length measure for a quantum gravity state is consistent
not only for the symmetries of the metric tenor, but also for the role its dependence on such symmetries has in the
series expansion.

The evolution of the metric tensor within the quantum phase of the universe can therefore be assumed therefore also different from the standard one, given that the conditions with respect to the Planck age be suitably determined according to the present observations: several mechanisms have been proposed, for which this condition can nevertheless be achieved.

In `Quantum Reduced Gravity', the symmetry group describing the metric tensor is interpreted as three independent
$U(1)$ fields, which then undergo suitable gauge fixing \cite{ales10}, and the connection variables internal ind-ices are therefore summed as
$\hat{A}_a=A_a\hat{I}$. The pertinent cosmological toy-model at the Planck scale and briefly
 after can offer a~simplification wrt LQC for the
definition of length measures. The corresponding quantum-gravitational problem allows one to understand its Planck-scale modifications \cite{Ashtekar:2013xka} at most as modifying the
free-particle dynamics as from the modifications of the kinetic-energy term in the Lagrangian: far from the singularity-
removal regions of `loopy' cosmologies
\cite{asht2006}, the picture can be at most that of a `frozen' anisotropic model.
This description of the qualifications of symmetries of the metric tensor below the Planck age is also consistent with the analysis of \cite{pots2000}, in which the states corresponding to more complicated symmetry groups decouple under the action of the Hamiltonian operator.

Differently, a slightly anisotropic solution is also possible by imposing small non-gravitational perturbations to an expanding (isotropic) FRW model, form an experimental point of view. Accordingly, the non-gravitational perturbations do not modify the gravitational potential at the proper order. Furthermore, the perturbation for the isotropic model can be tailored such as to produce the same affects of different initial conditions for the Einstein filed equations of an anisotropic model.

\subsection{Semiclassical Gravitational Functionals}
Quantum gravity condensates are solutions to a second quantization field theories which reproduce the degrees of freedom of quantum gravity on a manifold, without being, nevertheless, the non-flat manifold not being nevertheless quantum spacetime. A homogenous description of spacetime and its non-homogenous version are of a quantum theories of discrete geometries which lead to large-scale observables \cite{Gielen:2014ila}.

The expression of the degrees of freedom provided by quantum gravity condensates is also consistent with the description of spacetime as collective states of matter \cite{Hu:2005ub}.
Classes of quantum-gravity condensates which give rise to well-defined macroscopic geometries are analyzed in~\cite{Gielen:2013kla}.
Correction to the CMB for the corresponding semiclassical sate of the universe are calculated in~\cite{Kiefer:2011cc}, where upper bounds on the energy scale of the inflationary process are posed.

The passage from quantum geometrodynamics to quantum theory on fixed background is viewed in \cite{Kiefer:2008bs}, where the suppression of quantum features of the modern universe is assured, and the (almost) linearity in the probabilities is discussed in \cite{Mueller:2016aov}, by the phenomenon of decoherence. The corresponding General-relativity model without a fixed metric has been proposed in \cite{Ashtekar:1994wa}.
The role of quantum instabilities within the passage is further investigated in \cite{Calzetta:1993qe}. Dissipation for fluctuations and the pertinent noise are examined in \cite{Hu:1994dka}.
\subsection{Semiclassical States from Quantum Cosmologies}
The wavefunctions for semiclassical states of quantum theories of the spacetime, i.e., those possible to be described at the Planck length-scale, are different for quantum-gravity models and for quantum-cosmology models \cite{ashtx}.\\
The possible choices for the initial conditions in quantum-gravitational models \cite{incond1} and their relevance within the evolution of the quantum phase of the universe have been, in addition, compared \cite{incond2} wrt the possibility to match the boundary conditions at the semiclassical phase transitions. In \cite{borde}, the complementary problem is analyzed without assuming the effects of the quantum features of spacetime as important as to modify the information contained in the metric tensor.

The generation of non-Gaussianities \cite{Agullo:2012sh} can be included within this framework by taking into account the possibility of simplifying assumptions for the dynamics of the anisotropic universe.

The importance of the choice of the time scale (according to the Planck mass \cite{anto}) with respect to the possible \cite{Rovelli:2005yj} importance of primordial gravitational radiation in the comparison of short-distance gravity (at present times) is outlined in \cite{gasper}, where the two different cut-off's needed are outlined; in particular, the latter is expressed as a vacuum-fluctuation (quantum) effect.

A similar investigation guideline is not a priori modified for the features of classical(ized) gravity ({\itshape{emerging}}) after the Planck scale, as defined by the corresponding (possibly) isotropic universe volume~\cite{jacob}. Also within this perspective, the effects of primordial perturbations can overlap those remaining from the anisotropic cosmological model, in the sense that a minimal length, which defines the volume (age) under which the universe is not explorable, can modify the Friedmann equation suitably also for generating inflation \cite{moham}.

\subsection{Semiclassical Wavefunctions}
Among several phenomena possible in quantum cosmology \cite{Grishchuk:1993ds} at the time of classicalization, non.vacuum solutions can be analyzed according to their properties of allowing for the description of space anisotropies whose features are still observable at present times. 
A model for a discrete perturbation spectrum is proposed in \cite{Hogan:2003mq}: a self interaction between quantum states in standard quantum field theory should therefore correspond to interference-like phenomena in the semiclassical quantum-mechanical wavefunction of the universe.

A description for such states consists of \cite{Hogan:2005iw} quantum-mechanical wavefunctions with spatially-bounded correlation functions; the corresponding inflationary model for an isotropic universe predicts space correlations logarithmically distant (with respect to the phase-space expansion). The same procedure does not apply to a quasi-isotropic model, for which \cite{Lesgourgues:1996jc} a coarse-graining (discretization) procedure is required for the analysis of the spacings of the space correlations. The~conditions which do not modify the isotropic volume expansion of the universe are analyzed in~\cite{Polarski:1995jg}. 
\subsection{Classicalization at the Hubble Radius}
The classicalizing effects of a scalar field within the framework \cite{Grain:2017dqa} of a slow-roll regime for the inflaton field assure to integrate out spectator fields \cite{Enqvist:2012xn}, for ages shorter than the inflation age.

The present observational indistinguishibility of quantum fluctuations from classical ones is connected \cite{Kiefer:2008ku} with the assumed value of the spatial curvature of the universe models. The effects of inflation on the wavefunctional of the universe \cite{Nelson:2016kjm} can be studied by the minimal coupling of a volume element to gravity. The comparison of the loss of information after the classicalizing transition is also compared to quantum effects of the gravitational interaction. Differently, such quantum effects are calculated to be negligible \cite{Burgess:2006jn} for the case of primordial fluctuations, according to their time definition with respect to the Hubble scale. The time of decoherence can be estimated according to the (density of) numbers of modes exceeding the Hubble radius, while the presence of inhomogeneities is shown to be linked more to the presence on initial quantum fluctuations rather than to the matter-system evolution interactions \cite{Calzetta:1995ys}.

	Discrepancies between the quantum dispersion of the semiclassical universe and the present observations can be analyzed as descending
	\cite{Martin:2015qta} from the compensation between the decayment of the two-point correlation functions, which should compensate for the presence of tensor gravitational degrees of freedom at the quantum level. Such degrees of freedom can be also explained \cite{Nelson:2016kjm} to produce small fluctuation on the wave functional, which however do not produce anisotropies but only a~decohered wavefunction, and \cite{Markkanen:2016jhg} lead to the observed (almost flat) curvature of the universe. The~quantum-mechanical fluctuations can also account for \cite{Boyanovsky:2015tba} short-wavelength degrees of freedom in the long-wavelength fluctuation analysis in the proper Hamiltonian description. Within the Hamiltonian description, the quantum-gravitationally- motivated effects on the gravitational wavefunction(al) can be calculated to have a spread in the spacing of the energy levels proportional to the (squared) frequencies \cite{Gambini:2004bm}, i.e., $\omega^2\sim T_{
Planck}^{4/3}/T^{2/3}$, and imposed a fully-covariant cut-off \cite{Chatwin-Davies:2016byj}.

\section{Flat Present Quantum Gravity\label{section5}}The sole uncertainty principle for an almost flat ($3+1$-dimensional) spacetime has been related, in~\cite{hogan}, within its holographic description, with the information loss for the ratio of the different (angular) orientations of null trajectories, for which case the flat-spacetime limit allows for the comparison. This~is consistent for the hypothesis of gravitational contribution(s) of the gravitational interaction to the correlation functions of macroscopic (gravitational) states \cite{macr2017}.

The possible detection of phenomena due to Planck scale quantum geometry effects is described in \cite{hogan4}, in which the difference between a (n also heuristic) modification of the metric tensor is calculated. The search for the difference with a fluctuating metric tensor is motivated by the need to neglect self-interaction for gravitons at experimental distances (i.e., those of apparati consisting of interferometers), for which a (resulting) effective quantum mechanics is consistent.\
The difference between the measure of length in a quantum-gravitational setting and in a quantum-gravity model is therefore the difference in the theoretical interpretation (and mathematical definition) of the eigenvalue corresponding to the square of a length (operator) and their eigenvalues \cite{puzio}.

In \cite{hogan4}, the ideal possibility of preparing an interferometer experiment on newly-emerged (quasi) Minkowski flat space time is envisaged. The minimum angular uncertainty $\Delta\Theta_{x_i}$ for the (newly-emerged geodesics corresponding to a photon) trajectory \cite{hogan3} (in the newly-emergent quasi Minkowski flat spacetime) is considered equivalent in all the coordinate directions for the interferometer arms of length $L$, $\Delta\Theta$, and limited as
\begin{equation}\label{deltatheta}
\Delta\Theta>\sqrt{l_P/L}
\end{equation}
and corresponds to the uncertainty due to the emergence of spacetime at the (Minkowski-time synchronization) beginning of the trajectory. Differently, on an emergent non-Minkowski spacetime, the angle (operator) can be specified according to different directions \cite{Major:2011ry}; the spectrum of such an~operator therefore depends on the symmetries of the spacial part of the metric tensor.

The uncertainty principle is applied for  particle trajectories, i.e., within the wave-particle duality at the Planck scale, on this interpretative background in \cite{hogan2}; on the newly-emerged spacetime, four-dimensional deformations of General Relativity, such as those described by \cite{cardella}, can be applied, for which, while the action is still metric-independent, the first-order corrections do depend on the~metric.

Here, for the experimental setting  (The rest frame limit  
\begin{equation}\label{footn}
[\hat{x}_i, \hat{x}_j]=\frac{i}{2\sqrt{\pi}}l_P\epsilon_{ijk}\hat{x}_k
\end{equation}
allows for the definition of the square of the distance operator
\begin{equation}
\mid \hat{x}\mid^2\mid l\rangle=\frac{1}{4\pi}l(l+1)l_P^2\mid l\rangle\equiv L^2\mid l\rangle
\end{equation}), two different angular directions are separated by the distance eigenvalues for the direction eigenstates
\begin{equation}
N_{2S}=\sum_{l=1}^{l=l_R}(2l+1)=l_R(l_R+2)\simeq 4\pi\left(\frac{R}{l_P}\right)
\end{equation}
(upon which normalization is due). According to the experimental setting, the distance operators are defined according to the (different) (macroscopic) positions of the detectors.

The (resulting) uncertainty-principle contributions are therefore interpreted as holographic noise in theories of emergent gravity \cite{emergent}.

The possible present-time deviation form classical gravity are explored by measuring the possible shear symmetry effects between orthogonal directions among all those selected in the interferometer apparatus for (orthonormal) macroscopic states (directions) \cite{hogan5}. The detection of a shear symmetry not only allows for the definition of a correlation function, but also for the discrimination between classical gravity (the almost-flat spacetime limit of General Relativity) and other non-commutative/relative-locality models.

The time derivative of the correlation function, necessarily of order $c^2t_P$, can also be estimated by further data analysis of the same experimental apparatus \cite{hogan6}. The time symmetry of the correlation functions allows one to discriminate between General relativity effects and cosmological signals. The correlation function $\Xi=L\xi_0$ needs therefore be normalized according to the length of the interferometer arm $L$, and so its time derivative $\mid\dot{\Xi}\mid=c\xi_0$.

This description of quantum mechanics at interferometric scales is \cite{hogan9} therefore completed by the definition of a length operator
\begin{equation}\label{geodesics}
\hat{X}(t)=\hat{x_2}(t)-\hat{x_1}(t-2L/c),
\end{equation}
i.e., such that the commutator of two such operators must forcedly be a trigonometric function of the coordinates times a constant bringing the Planck length. The wavepacket description for the (position) states are a function of the group velocity and on the transverse wavenumber, according to the direction (beam-splitter reflection). Interaction with matter is described by the appearance of a~transverse velocity (wavenumber). The phase-difference observable has a variance that depends on the position(s) of the states; the minimum of this variance corresponds to the standard deviation for the position observable, $\Delta X=\sqrt{2Ll'_P}$, $l'_P$ a length close to the minimum observable noise, the maximum wavenumber being inversely proportional to it, and $L$ the interferometer length. The correlation function at a (beginning of the experiment) proper time is proportional to the total variance, while the auto-correlation function to its (time) evolution, The time domain for the correlation function also fixes its spectrum.

The observed (statistical) noise is therefore to be connected with metric fluctuations, possibly produced at the Planck length; their behavior is dependent on frequency \cite{hogan10}.

Also in this case, the `complementary' problem has been checked in \cite{alwis}: the condition of the pre-inflationary phase can be resolved, as far as experimental observations are concerned, by imposing a suitable (Gaussian) distribution in the Kernel for the gravitational field, deriving from general assumptions \cite{bunch}. The observational implications are probatorily proposed by substituting the Planck mass by (some) other unit: this allows for the comparison with the experimental implications of other quantum theories of gravity \cite{wein}.

\subsection*{Length Errors}
The error in length measurements due to both quantum mechanical effects and relativistic ones for the components of the metric tensor is calculated in \cite{Ng:1994zk} as
\begin{equation}\label{ngz}
\delta g_{\mu\nu}>\simeq (l_P/l)^{2/3}>\simeq (t_P/t)^{2/3}.
\end{equation}
 A lower bound on the quantum mechanical contribution to the total error can be estimated to be local, and, in particular \cite{Ng:1993jb,Salecker:1957be,AmelinoCamelia:1994vs}, to depend on the inverse of the initial position $x_i$ of the measurement system, i.e., 
\begin{equation}\label{disgu1}
\delta x\ge \delta x_i+\frac{h}{c}\frac{2L}{m\delta x_i}.
\end{equation}

Quantum-gravitational models can allow an intuition of the Planck-scale description of the spacetime as a most general expansion of the white noise as as powers of the Planck length \cite{amel}, whose peculiarities can account for those a particular quantum-gravity model. Such a power spectrum of the frequency $\varphi$ of the strain noise, according to its most general features for the detection of `foamy' features of the spacetime, needs be
\begin{equation}\label{needsbe}
\rho\simeq\sum_{n\ge0}a_n\left(\frac{L_P}{c}\right)\varphi^n
\end{equation}
with $a_n$ numerical coefficients; inverse powers need not be included, as they are inconsistent with the theoretical classical limit $L_P\rightarrow0$ \cite{amel2}.

The total uncertainty of an experimental measure having to depend on the sum of the uncertainty of the initial and final position, which cannot be smaller than the eigenvalue spacing (i.e., the Planck length) \cite{calmet}.

Modifications about the  errors in measuring lengths can be obtained by considering the quantum nature of extended bodies instead of their macroscopic ruler feature \cite{Requardt:2008yy}. The gravitational nature of macroscopic clock(- system)s is taken into account in \cite{Adler:1999if}, where they are commented to be expected not to exhibit quantum properties much above the Planck scale. The quantum nature of rulers of mass smaller than the Schwarzschild limit is expected \cite{Baez:2002ra} to be manifest at distances comparable to the Planck scale or less than one order of magnitude larger than it. The Planck-scale fluctuations (for vacuum) in measurement apparati \cite{Requardt:2005yw} and the space range of the vacuum fluctuations in the apparati has been analyzed to be also possibly large, but subject to fine tuning.
The physical passage form quantum foam to General Relativity is presented in \cite{Cahill:2002gs}
\section{Observational Tests\label{section6}}
\subsubsection{Hilbert space} The Hilbert space prepared for lattice gauge theory is analyzed in \cite{gran2}. Here, the eigenstates of the Hamiltonian operator are evaluates in the flat spacetime limit and in the strong field regime; the~vacuum state is proportional to the zero-value eigenvectors of the Hamiltonian operator, and the efficiency of the numerical methods is compared to the analytical ones.

The small-scale structure of a spacetime are reviewed in \cite{gibbs1}; the difficulties on the definition of areas and volumes within the quantum framework are analyzed in \cite{gibbs2}.

	The possible phenomenological effects of the existence of a pre-geometry are outlined in \cite{pregeo2}; the links between causality and discretized geometry \cite{Henson:2006kf} are discussed in \cite{Bombelli:1989mu} also as far as Lorentz violation is concerned.
	\subsubsection{Geodesic modifications to quantum dispersion relations}
	Any modification to the geodesics equation arising from non-commutative \cite{cardella} gravity can also be applied to the Planckian setting. Pregeometry non-commutativity scenarios before the Planck time can be described as first-order $h$ in the geodesics equation \cite{Pramanik:2014mda}. A Lorentz-violating pregeometry also can imply a modification at the same order.\\
	
	After upgrading lengths to operators, for Lorentz-violating pregeometries, their expectation value is supposed to undergo parity description for states \cite{Sorkin:2011sp}

	containing information about the metric tensor $\Psi_{\Gamma}$; in the case such metric tensor is one allowing an evolution to a (n, also, perturbed) FRW classical(ized) metric, it can be demonstrated \cite{Kempf:2012sg}

	that expectation values at first order in the Planck length are highly suppressed after the Planckian time, such that the considered states are those whose operators expectation-values are second order in the Planck scale.\\
	\\
	The phenomenologically most modifying description, as far as Minkowskian spacetime is concerned, is that of \cite{Pramanik:2014mda} classicalizing non-massless gravity. For this, at Planckian scale, the dispersion relations depend on the dimension of the operator for which they are evaluated,
	\cite{Mewes:2013cda},

	and read $k^2=\omega^2(1\pm((M\omega)/(M_P^2))^{N-4})$, $M$ being the phenomenological parameter accounting for massive gravity, and $N$ such dimension. After the Planckian scale, its expression for length measurements $k_L$, as a (Taylor) expansion, is well-defined, i.e. also consistent with a phenomenological agreement, as it can be rewritten as
	\begin{equation}
	k_L=\omega_L\sim \omega+\mathcal{O}\frac{3}{2}\left(\frac{\omega}{M_P}\right)^{-3/2}
	\end{equation}
	in the classical gravity case, where the dimension of the length operator $L$ is therefore $N=1$.\\
	The limit for the dispersion relations is therefore well-defined, and its effect is to modify the frequency $\omega$ according to the semiclassical limit $\omega_L$ of the quantum-gravitational length operator for which it is evaluated at the Planck size.\\
	\\
	The understanding of the effects of Lorentz-violating (pre-)geometrical settings, which can be described as classicalizing under the same (upper-bound) constraints for the evolution of semiclassical perturbed FRW states $\Psi_{\Gamma}$, any such description will lead to the analysis of different
	\cite{Shankaranarayanan:2004dk}

phenomenological tracking in the description of the background temperature fluctuation.\\
The description of the need for experimental apparati and measure instruments needed for measurements of th leftovers of Lorentz-violating pregeometries on the parity-conservation of the classicalized pregeometry states detectable on cosmological-distance-scales experiments \cite{Kaufman:2014rpa} has been outlined after both the

	theoretical possibility \cite{Halpern:1999ng} and the phenomenology statistics \cite{HernandezMonteagudo:2002df} analysis

	has been performed.
\subsection{Geometric Quantum Tests} 
The consequences of quantum spacetime granularity are described in \cite{gran1}, as violations of the Lorentz symmetry and Planck-scale modified electrodynamics. The role payed by the Lorentz symmetry below the Planck scale and its 'quantum' effects at the Planck scale for the gravitational field are investigated in \cite{gran5}. The phenomenological limits of the resulting macroscopic Einsteinian gravity are analyzed in \cite{gran3}, while those in a uniform Newtonian field in \cite{gran4}.

	The need for a pre-geometric description of the spacetime, within its quantum formulation, i.e.,~below the Planck age, is discussed in \cite{pregeo1}.
	
	The possible violations of Lorentz symmetry are explained to effect the propagation of photons and of neutrinos at astrophysical scale \cite{sark}. Lorentz violations at the Planck length are shown to produce modifications in the dispersion relations, for which the interaction of high-energy particles at astrophysical scales can be expected to be different \cite{sark2}, for which effective experimental uncertainty windows can be considered.
	
	 The (field theoretical) dimension of operators affected by Lorentz symmetry violations are analyzed according to the particle on which the effects are detectable at astrophysical scales \cite{sark3,fant2014a}.
	 
The phenomenological possibility to further find such effects is hinted in \cite{Girelli:2012ju} as far as CMB and gravitational waves are concerned.

The stages of the universe evolution according to this model found after the cosmological singularity time and before the inflation time can acquire new non-Gaussianities \cite{Agullo:2013ai}.	
	Treating matter perturbations and metric perturbations differently within this framework has been pointed out to lead to observable effects in the present CMB spectrum \cite{Wilson-Ewing:2016yan}.\\
Perturbations of a Minkowski spacetime can be solved as resulting of two perturbed vacuum states, which are described by a characteristic amplitude. The characteristic amplitude, normalized at the present-day Hubble frequence \cite{grish}, is independent of frequencies for the deSitter case.
The definition of the corresponding data is produced by the superposition of several effects of different origin, which have not been fully understood yet, but modelized according to the needed distributions.\\

	\subsection{Sky Pixellization}
	Distance-measurements analysis at astrophysical scales is also involved in the identification of the origin of possible General Relativity violating phenomena, for which the implications on the components of the metric tensor can be not only described, but also ascribed to different quantum-gravitational models.\\
		\\
The necessary ratio of the two length measures is also consistent with the analysis of the correlation function  \cite{Sung:2010ek,chall}  for the
corresponding initial conditions at the given time to the most general investigation of the temperature 
fluctuations
\begin{equation}\label{cmb}
\Delta T_{\theta, \phi}=\frac{T_{\theta, \phi}-\bar{T}}{\bar{T}}=\sum_{l=0}^{l=\infty}\sum_{m=-l}^{m=l}\mid a_{l, m}\mid e^{i\Phi_{l, m}}Y_{l, m}(\theta, \Phi)
\end{equation}
where $a_{l, m} $ are the amplitudes, $\Phi_{l, m}$ are the phases, $Y_{l, m}$ the spherical harmonics coefficients, where,
for Bianchi geometries, the amplitudes are not stochastically generated \cite{king2006,rockhee2010}, while the distribution of the phases can be
generated according to different methods.

A discrimination with respect to the sampling of the astrophysical foreground for the CMB analysis in the definition of distances between Sky patches is therefore possible. In particular, the expected (superposed) error is understood to be consistent as in the expansion (\ref{needsbe}) also without the hypotheses of quantum-gravitational theories. The corresponding error(s) for the components of the metric tensor are therefore described as perturbations of General Relativity within the due observational uncertainty, with $a_n=a_n(\varrho)$ from (\ref{ngz}) in (\ref{needsbe}).\\
For a more precise setting, i.e., after the indication \cite{Cho:2003af}  for the need of more precise identifications of the components of the CMB experimental data \cite{ichik14},

the astrophysical foreground patterns have been outlined \cite{Adam:2015tpy} \cite{Adam:2015wua} \cite{Ade:2015qkp} and removed \cite{deOliveiraCosta:2002ng}
for fulfilling the attempts to
   theoretically recognize \cite{Choi:2015xha} and distinguish a (possible) gravitational origin \cite{Cho:2010kw} for the components of the harmonic decomposition of the Sky pixellization patches \cite{Yamauchi:2016ypt} \cite{Ruud:2015pvs}	\cite{Namikawa:2015tba}.

The differences of the temperatures samples can indeed be analyzed with the aim to investigate the implications of classicalization of perturbations as quantum-gravitational contributions at the Planck epoch. To do so, the Sky temperature patterns have to be reconducted to the proper cosmological analysis interpretation by the
\cite{Rockhee2009} right choice of the direction axes needed by different Bianchi models. The analysis of small perturbations to the isotropic growth of an FRW universe is therefore reconducted to the analysis of a some Bianchi configuration with given initial condition for the Einstein field equations. The topology for (selected) Bianchi universes has been investigated in \cite{planck2013}.

Non-Gaussianities are expected to be outlined within \cite{Purkayastha:2017sjj} the astrophysics (i.e. astronomic-techniques-based) harmonics decomposition

\cite{diSeregoAlighieri:2016lbr}

	; within this framework, the different origin of the phenomena can be distinguished according to the tracking of the metric tensor.\\
	\\
	As an example, the astrophysical foreground \cite{Zhai:2017ibd}

	has been analyzed to be consistent with a mode decomposition comparable with the analysis of the Sky patches \cite{Gorski:2004by}\cite{Bucher:2015eia}.\\
	\\
Because the requested small perturbations do not modify the gravitational potential, it is not possible, within this experimental point of view \cite{planck2015}, to discriminate between gravitational perturbations and non-gravitational ones \cite{fant2014}. As such phenomena are supposed to have happened at the Planck age within a semiclassical regime, the assumption of quantum-gravitational phenomena is therefore allowed.

\section{Length Observations from Quantum Cosmology\label{section7}}
Modification for General Relativity can be measured today according to two different origins, i.e., either they are encoded in the metric tensor as the solution to Einstein field equations, or they are quantum modifications of the Minkowski spacetime. 
Any (observed) modification of the isotropy of the Universe must therefore be described as non-second-order measured
length ratios
\begin{equation}\label{geodesi}
\frac{\langle \hat{L}(\gamma_{x_a}(t_2-t_1))\rangle}{\langle\hat{L}(\gamma_{x_b}(t_2-t_1))\rangle}\sim 1+\epsilon(f_a-f_b)+\mathcal{O}\left(\frac{l_{Pl}}{\Lambda}\right)
\end{equation}
for segments of length 
$\gamma_{x_a}<\Lambda$ evaluated along two of the three space directions $\gamma_{x_a}<\Lambda$ evaluated along two of the three space (coordinate) directions $x_a, x_b=x, y, z$ in the time interval $t_2-t_1$,\\
with $\pm\epsilon$ \textasciitilde1 in the case both curves are evaluated on the Planck length. The request $\pm\epsilon$ \textasciitilde1 englobes the normalization with respect to the length $\hat{L}(\gamma_{x_b}(t_2-t_1))$ in the series expansion at the Planck length. More in particular, it is supposed also to factor out any proper time dependence of the length measure.

Because the expectation values are non-vanishing, a semiclassical state $\Psi_{\Gamma, c}$ contains information on the metric
tensor;
because of the non-trivial ratios observed, a semiclassical state is characterized 
 by the initial conditions for the Einstein field equations which imply a specific space three-geometry.

\subsubsection*{Hamiltonian Limit}
In particular, any dependence on quantum-gravity effects $f_i$ in (\ref{geodesi}) of semiclassical states must be such that the quantities $f_i(\vec{x}; \beta)$ become
\begin{equation}\label{bila}
\lim_{\beta\rightarrow0}f_i(\vec{x}; \beta)=f_i(\mid\vec{x}\mid; x_i),
\end{equation} 
where the normalization $\mid\vec{x}\mid$ is performed with respect to the metric tensor.

This is such that the modification of the time evolution (given by the Hamiltonian constraint), which also modifies the composition of the diffeomorphism constraint (under which General-Relativity lengths must be invariant) has the semiclassical contribution of implying differences in the choice of space directions, whose (semi-)classical limit has been achieved at least logarithmically, independently of the number of spacetime dimensions, as implied for the heat-kernel expansion, and has therefore~classicalized.

The choice of the limit in (\ref{bila}) allows one to concile any fine-tuning requested (or imposable) on the prefactor $\epsilon$ in (\ref{geodesi}) with the Hamiltonian formulation of (\ref{elemlen}).

Differently understood, this analysis also renders explicit the need that any such definition be of gravity origin, i.e., related to the metric tensor, and not related to other (external) phenomena.

From a quantum point of view, the case of perturbations to the 'heuristic flat' Minkowski spacetime, the correction limit (\ref{bila}) can therefore consist (also) as the contribution $\mathcal{O}\left(\frac{l_{Pl}}{\Lambda}\right)$ in (\ref{pulli}), as (\ref{bila}) has the correct corresponding Hamiltonian (formulation) limit.
\section{Comparison and Discussion\label{section8}} 
So far, it is possible to conclude that any anisotropy feature of the metric tensor calculated as from the ratio between lengths in two different (coordinate) directions should be distinguished, between the experimental error, from the same comparison applied to the investigation of quantum-gravitational (spinfoamy) effects (\ref{geodesics}) exhibited by (quasi) Minkowsky flat spacetime $\hat{X}(t)/\hat{Y}(t)$,
\begin{equation}
\frac{\hat{X}(t)}{\hat{Y}(t)}\simeq 1-\left(\frac{\hat{x}_1(t-2L/c)-\hat{x}_2(t-2L/c)}{\hat{Y}(t)}\right),
\end{equation}  
i.e., where the synchronization of measurements is performed according to Special Relativity, for which the difference in definition is of gravitational origin, i.e., according to the (non-Minkowski) metric tensor used for (\ref{geodesi}).

This discrimination can be applied also for the (normalized) background temperature fluctuations in (\ref{cmb}) by comparing the temperature differences according to the requested (Sky) patches~(pixellization).

According to the different patches, i.e., by naming (the center of) each patch according the radial coordinates $\theta_\alpha, \phi_\alpha$ and so on, the ratio (\ref{geodesi}) defines, up to a normalization in observational distance $X$,
\begin{equation}\label{varro}
\frac{X_\alpha}{X_\beta}\simeq 1-\varrho\frac{F_\alpha-F_\beta}{X_\beta}
\end{equation}
as functions of the $F$'s, i.e., the (correspondingly-normalized) differences in (\ref{cmb}), with $\varrho$ a numerical factor with the dimensions of [$length$].

This procedure explains that, after any further finetuning for the parameter $\epsilon$ in (\ref{geodesi}), according to the non-isotropic features modifying the standard cosmological thermal history of the universe, the modification term in (\ref{geodesi}) must be molded according to the present measured values. This is consistent with the requirements that the harmonics coefficients in the harmonic expansion (\ref{cmb}) encode the features of the metric tensor.\\
\\
Interestingly, the possibility of non-gravitational perturbations acting on a homogenous universe is by this experimental proposal allowed by the corresponding theoretical analysis. Furthermore, at~quantum ages, even an empty universe is supposed to exhibit quantum properties (at least also from a~thought experiment). Nevertheless, this possibility has to be ruled out, as it does not comprehend the presence of matter.\\
\\


In a cosmological setting, the expansion (\ref{geodesi}) is understood as normalized to the Plank length only if one coordinate direction, $b$, is integrated over a segment of the space volume growth of the universe e, whose length is proportional to the Planck Length. In this case, the prefactors $\epsilon$ is further factorized as $\epsilon\equiv\varepsilon_1(t)\varepsilon_2/\varepsilon_3(L_P)$, where $\varepsilon_3(L_P)\simeq L_P$, and $\varepsilon_2$ a numerical factor, whose dimensional analysis must agree with the normalization of the cylindrical functions. Furthermore, the $\varepsilon_2$ prefactor should also depend on the other (coordinate) direction, such that the coordinate directions are understood as the physical degrees of freedom of the system, and are understood as defining the metric tensor.

The expansion (\ref{geodesi}) therefore compares with (\ref{needsbe}) only in the weak-filed limit, i.e., only on those cases for which (\ref{geodesi}) corresponds to the Taylor expansion with respect to the Planck Length $L_P$, in~which it can be continued at higher orders. Its Hamiltonian formulation is correct, and should match a~considered pregeometry by being imposed the appropriate (boundary) conditions.

\subsection{Discussion}

According to the analysis in the above, it is possible to find in a length measured at present times a possible witness for deviations from a perturbed FRW metric as a modification for the corresponding metric tensor.

According to the most general paradigm, such deviations must be found within the experimental error, for which the uncertainty in not second order. More in particular, a measured length $L$ for a~(coordinate) length $L_{coord}$ form (\ref{quasim}) should correspond to
\begin{equation}\label{disgu}
L=\int_0^1\left(q_{ab}(c(t))\dot{c}^a(t)\dot{c}^b(t)\right)^{1/2}d^4x\sim L_{coord}+\Delta\Theta+f(g)
\end{equation}
where any fluctuation related with an emerging Minkowski spacetime is related to (\ref{deltatheta}), and any further non-Minkowski contribution is encoded as a function of the metric tensor $g$ from (\ref{bila}), as from the anisotropic-metric correction (\ref{geodesi}) (as focused in \cite{cardella}).

The comparison with (\ref{needsbe}) results as (\ref{geodesi}) by looking for the contributions of (\ref{geodesi}) in (\ref{deltatheta}).

Applying the analysis of (\ref{needsbe}) to the contribution of (\ref{geodesi}), the found contribution consists, at this limit, as a \textit{numerical} contribution to the first-order terms in (\ref{needsbe}), and does \textit{not} consist of a fine-tuning.
\subsection{Comparison of Modifications}
In the present sections, the modifications to the length of geodesics in the case of quantum effects from the Minkowskian scheme and those due to anisotropies of early-cosmological origin are~compared.

In particular, it is possible to establish the differences both in the theoretical schematization and in the experimental analysis for the components of the metric tensor $g_{0i}$ and $g_{ij}$.

In the case of emerging flat Minkowski spacetime, the contribution to different length measures can be ascribed to a non-vanishing values of the components $g_{0i}$ of the metric tensor, $i=1, 2, 3$, modifying the Minkowski metric in the line element. In particular, it is sufficient for the experimental difference to be noticed only one (opportune) value $i$, the other $j$ and $k$ such that $g_{0j}=g_{0k}=0$, the~difference arising being nevertheless of one order in $c$ smaller. This modification of can be therefore compared with those of the kind in (\ref{disgu1}), which are not due to the angular indeterminacy if the length measure, but comparable at the same $\mathcal{O}(1/c)$ order or smaller; indeed, in the case of further parametrization, i.e., such as $g_{0i}\simeq \mathcal{E} g_{0i}$, further fine-tunings are possible to shift the order of the correction, i.e., such that, with $\mathcal{E}\sim\mathcal{O}(1/c)$, $f(g)\simeq\mathcal{O}(1/c^2)$ in (\ref{disgu}).

This also compares possible foamy features of the spacetime, where any foamy feature results as `internal' (geometric) degree of freedom, which can be sampled at the pertinent $c$ order.

Modifications to an isotropic FRW model due to anisotropies are described by a symmetry with $g_{ij}\neq0$ for some $i\neq j$. In this case, the inhomogeneities can be interpreted as due to the presence of (microscopic) matter, and for which the small order of the contribution is due to the (small) mass of the matter present.

In the inhomogeneous case, such modifications are evaluated as `matter presence' without modifying the Ricci scalar, such that the corresponding
analysis is the same as that for the line element emerging modified Minkowski gravity in the above.

\subsubsection{Classicalized solutions}
At Planckian times, a modification to the general cosmological solution can be considered for the calculation of the line element, such that the contributions originating after modifications to the scale factors are majorized, at second order, by $1/l_P$.
A modification of the geodesics lengths measuring coordinate intervals observed at (semi)-classicalization Planckian time can be compared, in the most evident modification, to a running feature of the gravitational constant $G$, such that $\frac{1}{G}\rightarrow\frac{1}{G}+\frac{1}{\mathcal{G}}$, $\mathcal{G}\equiv\mathcal{G}(x_i; \mid \vec{x} \mid)$ for the analysis at/after Planckian times.

The corrections consistent with the case of a quantum-gravitational correction of the gravitational constant due to two masses are described by a modification term quadratic in distances \cite{Anber:2011ut,BjerrumBohr:2002kt}, for~which $\mathcal{G}=r^2$.

A different choice majorizes this result, i.e.
 $\frac{1}{\mathcal{G}}=\frac{G}{\mathcal{G}(x_i; \mid \vec{x} \mid)}$, and, in particular,
\begin{equation}
\frac{1}{G}\rightarrow\frac{1}{G}\left(1+\frac{\mathcal{C}}{\pi}\frac{1}{r}\right),
\end{equation}
with $\mathcal{C}$ a constant with the dimension of $[length]$; such an approximation therefore generalizes for quantum cosmological times the present asymptotically Minkowskian case.

If the corresponding Einstein equations are not solved by taking into account the corresponding modifications as due to a scalar field, and that by letting it (also, semi-)classicalize, the corresponding implications for anisotropy in the ratio of the two length measures is therefore majorized by the Planck length, such that the non-second-order correction in (\ref{geodesi}) results as $\epsilon(f_b-f_a)\leq l_P$.
The modification to the components of the metric tensor (\ref{ngz}) are therefore to be sampled out from the analysis of the astrophysical foreground. Such corrections are therefore expected to be as $\delta g_{\mu\nu}\equiv \delta g_{\mu\nu}(a_2(\epsilon), \varrho)< l_P$.  

Differently from the previous cases, in the case of a modification of the Einstein-Hilbert action, the~strongest effect due to the background temperature differences evidenced at early-cosmological times being depicted by, i.e., in the case of a Lagrangian density containing quadratic terms in the curvature scalar \cite{Khriplovich:2002bt}, such as $\frac{c^3}{16\pi G}(R+\frac{c^3}{16\pi G}q R^2)$, $q$ dimensionless, the difference arising for (\ref{bila}) is of second order in $(1/c)$, and consistent with quantum-gravitational modification on General Relativity~\cite{stelle1977}. 
In this case, the corresponding parameter $\varrho$ in \ref{varro} is expressed as $\varrho\equiv\varrho(q^2)$ (with respect to Special Relativity distances measurings or the definition of the Sky patches pixellization \cite{Conroy:2017nkc}). For~the same reason, an estimate of $q$ is therefore taken as letting Relativity be an opportune model for Sky observations data analysis at least for the Solar System, for which $q<<1/(r_{ss}/r_{\odot})$ (with \mbox{$r_{ss}$ $\simeq$ 10$\textsuperscript{11}$~Km} the radius of the Solar System, and $r_{\odot}$ the radius of the Sun $r_{\odot}\simeq 10^5$ Km. 

\section{Concluding Remarks\label{section9}}
Several proposals have been formulated for the description of the possible quantum properties assumed by the gravitational interaction at length scales below the Planckian scale. The corresponding motivations nevertheless must differentiate on the basis of whether the (observed) effects are described for cosmological models or on (thought) lab experiments. Indeed, the behavior of the gravitational interaction is different at regimes, according to the (pertinent expansion of the) Ricci scalar at the proper considered distances.

The gravitational interaction can be hypothesized to have underwent a quantum phase at cosmological ages below the Planck age.

The analysis of the classicalization of quantum-gravitational effects after the Planck age has to be modelized according to the possible experimental verifications of the presently-evolved models. The~analyzed paradigms should allow to look for the observable effects at the right order (in the expansion with respect to the proper length units).

The classicalized quantum effects here analyzed are assumed to be produced in their present aspect at the classicalization epoch. Such perturbations to a isotropic model fit all the requirements for perturbations to an FRW description. Such scheme does not specify whether the perturbations are gravitational or due to matter fields, the isotropic evolution of an FRW model with isotropically-distributed matter not being distinguished in this case.

Nevertheless, the presence of matter has to be taken into account in any realistic description of the matter distribution of the Universe.

The definition of the dependence of the result of length operators imposes the semiclassical wavefunctional to be well.posed with respect to the canonical Hamiltonian formulation of General Relativity, for which the limit to Special Relativity is well-defined with respect to any quantum-gravity effect, i.e., after the request that any quantum effect on length measures be diffeomorphism invariant. The invariance under diffeomorphisms of the pre-geometry for General Relativity has been discussed in
\cite{Prugovecki:1989kz,dre1,dre2,anandan}.

The analysis of the effects of the possible classicalization of quantum properties of the gravitational field has therefore been performed with respect to its implication on the evolution of the metric tensor after the Planck age, i.e., on the anisotropy directions individuated at the Planck scale, on which no possible modification paradigm has acted thereafter. The obtainment of this configuration for the metric tensor is compatible with a simplification of the standard cosmological model, for which strong anisotropies do not constitute a characterizing feature, an isotropization mechanism being needed. The same individuation of preferred directions is also compatible with small perturbations (of any nature) to an FRW model containing homogenous matter. All the possibilities have been outlined to be equivalent, as far as the observation of the present sky patterns depicts the empty universe.

Geodesic lengths are needed for the establishment of length (measures). The (natural) existence of preferred direction, i.e., the qualification of (three) preferred (coordinate) axes requires the assumption of the existence of solutions for the Einstein field equations, which are determined by the presence of matter even in /causing the strong field regime.

The analysis of the heat kernel for the gravitational field, which reproduces, at the given approximation (length expansion), parametrized geodesics, reveals the contribution of the curvature scalar, which is independent of the number of (space) dimensions adopted.

The presence of quantum effects for the gravitational interaction can be outlined, accordingly, by the presence of non-trivial killing vectors in an empty $1+1$ dimensional spacetime, for which the description of geodesics stays unchanged, as outlined by the heat-kernel expansion.

The very same result follows from the description in a dimensionally-reduced model, in which the necessary number of dimensions is $1+1^n$, $0\le n\le 1$ \cite{gac2015}, the structure of the corresponding killing vectors not being compatible with the presence of matter in a spherically-symmetric \mbox{$1+1$ model}~\cite{carlip0909}; it can be framed within the more comprehensive description of the difficulties to reconstruct a~classical four-dimensional universe from its spectral (dimensionally-reduced) analysis~\cite{Ambjorn:2005db}, for which deformations of General Relativity can be allowed.

Event though the possibility of quantum-gravitational effects at the classicalization epoch is compatible with the evolution of the metric tensor and therefore with the different temperature patterns observed, for which also the previous (classically needed) evolution of the metric tensor at quantum times is included, its results are not automatic.

A compatible prescription on the evolution of the so-happening celestial objects is therefore obtained. The actual distribution of such objects is nevertheless described by the presence of matter at the Planck age according to the evolution of the metric tensor. Its description does not enter the definition of such quantum-gravity effects. More in particular, such a description is compatible with the initial conditions for the Einstein field equations, and can therefore be analyzed according to the effects that such initial conditions have on the correlation function for a classicalizing universe.

In the present work, the features of the results of length measures have been outlined (\ref{geodesi}), such that a possibility to distinguish between the times at which modifications to the spherically-symmetric FRW model have been activated, i.e., whether those effects are still detectable at present-time experiments and distinguished from the possible (also  foamy) modifications to the quasi (Minkowski) flat spacetime. Such terms modifying the FRW metric tensor should be detectable as the parameters describing them should have stayed unchanged since the classicalization epoch, in which any quantum features of the gravitational interaction in vacuum should have stabilized on transPlanckian distances. This~is compatible with the limit of the semiclassical wavefunctional with respect to a Hamiltonian formulation~(\ref{bila}).

This is achieved by requesting that the ratio between two lengths in two different coordinate directions at the same proper time at the Planck age be factorized with respect to the time dependence in the Hamiltonian canonical formulation of General Relativity (\ref{geodesics}).

This most general description has been compared to the possibility to detect quantum-gravity phenomena due to the possibly intrinsic features of the gravitational interaction at present days, i.e.,~such that any diffeomorphism invariance is respected.

Such features of the metric tensor can also be compared with the present look of the Sky \cite{fant2014}, i.e.,~a~conceptual interest can be focused on the difference in temperatures (\ref{cmb}) in the sky pixellization with respect to the components of the metric tensor.


The paper has been organized as follows.

The delineation of the features of the gravitational field at within the quantum phase of the universe, i.e., the heat kernel for the gravitational field, are exposed in Section \ref{section2}.

The definitions of length operators effective both in the strong-field regime of gravity and in the quantum phases of gravity are analyzed in Section \ref{section3}.
 
The quantum-gravitational models that allow for a classicalization transition are reviewed in Section \ref{section4}.

The experimental evidences needed to infer quantum properties of quasi (Minkowski) flat spacetimes at present times are described in Section \ref{section5}.

The theoretical schemes that allow to look for evidence of weak anisotropies at the Planckian stages of the universe, such as the modelization of the analysis of the CMB radiation temperature fluctuations, are recalled in Section \ref{section6}. 
The comparison of different length measurements aimed at detecting classicalized quantum degrees of freedom is proposed in Section \ref{section7}, and compared with the previous analyses in Section \ref{section8}. In particular, the experimental description is related to the modifications to the components of the metric tensor.

\vspace{6pt}

\acknowledgments{
O.M.L. would like to thank Giorgio Immirzi for useful comments on the second paragraph of Section \ref{section6}.}  


\conflictsofinterest{Declare conflicts of interest or state ``The authors declare no conflict of interest''. ``The~founding sponsors encourage preparing manuscripts for publishing''.} 

\bibliographystyle{mdpi}

\begin{thebibliography}{999}
\bibitem{Lobo:2013vga} 
Lobo, F.N.S.; Martinez-Asencio, J.; Olmo, G.J.; Rubiera-Garcia, D.
Planck scale physics and topology change through an exactly solvable model.
{\it Phys. Lett.} \textbf{2014}, {\it B731}, 163--167. 
\bibitem{Ambjorn:2005db}
Ambjorn, J.; Jurkiewicz, J.; Loll, R.
Spectral dimension of the universe.
{\it Phys. Rev. Lett.} \textbf{2005}, {\it 95}, 171301.

\bibitem{carlip2009} 
Carlip, S.  The Small Scale Structure of Spacetime. In  \emph{Foundations of Space and Time: Reflections on Quantum Gravity};  Murugan, J.,  Weltman, A., Ellis, G.F.R., Eds.; Cambridge University Press: Cambridge, UK, 2013; ISBN:10  0521114403.



\bibitem{futa84} \textls[-15]{Futamase, T. Coleman-Weinberg symmetry breaking in an anisotropic spacetime. {\it Phys. Rev. D} {\bf 1984}, {\it 29}, 2783.}


\bibitem{Ambjorn:1998pz} 
Ambjorn, J.; Anagnostopoulos, K.M.; Ichihara, T.; Jensen, L.; Watabiki, Y.
Quantum geometry and diffusion.
{\it JHEP}  \textbf{1998}, {\it 9811}, 022, doi:10.1088/1126-6708/1998/11/022.
\bibitem{Zizzi:1999sx} 
Zizzi, P.A.
Holography, quantum geometry, and quantum information theory.
\emph{Entropy} {\bf 2000},  \emph{2}, 39--69.
\bibitem{Roh:2000na}
Roh, H.S.
Toward Quantum Gravity. 1. Newton Gravitation Constant, Cosmological Constant, and Classical Tests.
Available online: \url{gr-qc/0101001} (accessed on).





\bibitem{Major:2001ka} 
Major, S.A.; Setter, K.L.
Gravitational statistical mechanics: A Model.
{\it Class. Quantum Gravity} {\bf 2001},   \emph{18},   5125--5142.
\bibitem{momlab}
Torromé, R.G.; Letizia, M.; Liberati, S.
Phenomenology of effective geometries from quantum gravity.
{\it \mbox{Phys. Rev. D}} {\bf 2015},  \emph{92},  124021.
\bibitem{momlab2}
Assanioussi, M.; Dapor, A.; Lewandowski, J.
Rainbow metric from quantum gravity.
{\it Phys. Lett. B} {\bf 2015}, \emph{751}, 302, doi:10.1016/j.physletb.2015.10.043.
\bibitem{guth81}
Guth, A.H.  {\it Phys. Rev  D}  
\bfseries1981\mdseries, {\it 23}, 347.
\bibitem{lind83}
Linde, A.D.; {\it Phys. Lett. B}  \textbf{1983}, {\it 129}, 177. 
\bibitem{bianchi1997} Bianchi, E. The length operator in Loop Quantum Gravity.
{\it Nucl. Phys. B} \bfseries2009\mdseries, {\it 807}, 591--624. 
\bibitem{pulli99} Gambini R.; Pullin, J. Nonstandard optics from quantum spacetime.
{\it Phys. Rev.  D} \bfseries1999\mdseries, {\it 59}, 124021, doi:10.1103/PhysRevD.59.124021. 
\bibitem{Bombelli:2006nm} 
Bombelli, L.; Henson, J.; Sorkin, R.D.
Discreteness without symmetry breaking: A Theorem.
{\it Mod. Phys. Lett.~A}  {\bf 2009}, \emph{24}, 2579--2587.
\bibitem{Saravani:2014gza} 
Saravani, M.; Aslanbeigi, S.
On the Causal Set-Continuum Correspondence.
{\it Class. Quantum Gravity} {\bf 2014}, \emph{31}, 205013, doi:10.1088/0264-9381/31/20/205013.
\bibitem{Hedrich:2009pb} 
Hedrich, R.
Quantum Gravity: Motivations and Alternatives.
\emph{Phys. Phil.} {\bf 2010}, \emph{2010}, 016.

   
   
   
\bibitem{quantumgraphs}
Kuchment, P. Quantum graphs: An Introduction and a brief survey. 
In {\it Analysis on Graphs and its Applications},  \emph{Proceedings of Symposia in Pure Mathematics}; AMS: Providence, RI, USA,  2008; pp.~291--314.



\bibitem{loc}
Dragoman, D. Special Relativity in Quantum Phase Space.   \emph{arXiv}, arXiv:0803.0972 [quant-ph].




\bibitem{Krugly:2011np} 
Krugly, A.L.; Stepanian, I.V.
An example of the stochastic dynamics of a causal set.
\emph{AIP Conf.   Proc.} {\bf 2012},   \emph{1424},  206--210.
\bibitem{Rovelli:2006fw} 
Rovelli, C.; Speziale, S.
A Semiclassical tetrahedron.
{\it Class. Quantum Gravity}  \textbf{2006}, {\it 23}, 5861. 
\bibitem{Livine:2006ab} 
Livine, E.R.; Speziale, S.; Willis, J.L.
Towards the graviton from spinfoams: Higher order corrections in the 3-D toy model.
{\it Phys. Rev. D} \textbf{2007}, {\it 75}, 024038, doi:10.1103/PhysRevD.75.024038.
\bibitem{Rovelli:2005yj} 
Rovelli, C.
Graviton propagator from background-independent quantum gravity.
{\it Phys. Rev. Lett.} \textbf{2006}, {\it 97}, 151301,  doi:10.1103/PhysRevLett.97.151301.

\bibitem{area} Eisert, J.;  Cramer, M.; Plenio, M.B. 
Area laws for the entanglement entropy---A review. \emph{Rev. Mod. Phys.}  \bfseries2010\mdseries, \emph{82},  277--306. 
\bibitem{Hamma:2015xla} 
Hamma, A.; Hung, L.J.; Marciano, A.; Zhang, M.
Area Law from Loop Quantum Gravity. \emph{arXiv},  arXiv:1506.01623 [gr-qc].




\bibitem{single}
\textls[-20]{Chirco, G.; Rovelli, C.; Ruggiero, P.
Thermally correlated states in Loop Quantum Gravity.
{\it \mbox{Class. Quantum Gravity}}   \bfseries2015\mdseries,  {\it 32},  035011, 
doi:10.1088/0264-9381/32/3/035011.}
\bibitem{schlie1}
\textls[-20]{Schliemann, J.
Classical and Quantum Polyhedra.
{\it Phys. Rev.  D}   \bfseries2014\mdseries, {\it 90}, 124080,
doi:10.1103/PhysRevD.90.124080.}
\bibitem{schlie2}
Schliemann, J.
The Large-Volume Limit of a Quantum Tetrahedron is a Quantum Harmonic Oscillator.
{\it \mbox{Class. Quantum Gravity}} \bfseries2013\mdseries, {\it 30}, 235018, 
doi:10.1088/0264-9381/30/23/235018.
\bibitem{Bendjoudi:2016gom} 
Bendjoudi, A.; Mebarki, N.
The quantum tetrahedron and the length spectrum.
\emph{Int.\ J.\ Mod.\ Phys.\ D} \bfseries2016\mdseries, {\it 26},  1750044,
doi:10.1142/S0218271817500444.
\bibitem{bian1009}
Bianchi, E.; Dona, P.; Speziale, S.
Polyhedra in loop quantum gravity.
{\it Phys. Rev. D} \bfseries2011\mdseries, {\it 83}, 044035, 
doi:10.1103/PhysRevD.83.044035.
\bibitem{Loll:1995wt}
Loll, R.
The Volume operator in discretized quantum gravity.
{\it Phys. Rev.  Lett.}  \bfseries1995\mdseries, {\it 75}, 3048,
doi:10.1103/Phys\linebreak~RevLett.75.3048.
\bibitem{Thiemann:1996at} 
Thiemann, T.
A Length operator for canonical quantum gravity.
\emph{J. Math.  Phys.}  \textbf{1998}, {\it 39}, 3372--3392.
\bibitem{llft} Landau, L.D.; Lifshitz,  E.M.  {\it Classical Theory of Fields}, 4th ed.; Addison-Wesley:   New
York, NY, USA, 1975.
\bibitem{Loll:1996nk}
Loll, R.
A Real alternative to quantum gravity in loop space.
{\it Phys. Rev. D}   \bfseries1996\mdseries, {\it 54}, 5381,
doi:10.1103/Phys\linebreak~RevD.54.5381.
\bibitem{wong}
Ma, Y.; Soo, C.; Yang, J.
New length operator for loop quantum gravity.
{\it Phys. Rev. D} \bfseries2010\mdseries, {\it 81}, 124026, 
doi:10.1103/PhysRevD.81.124026.
\bibitem{Ma:2010fy}
Han,   H.; Huang,  W.;  Ma, Y.
Fundamental structure of loop quantum gravity.
{\it Int. J. Mod. Phys. D} \p2007\m, {\it 16} 1397--1474.
\bibitem{ansari}
Ansari, M.H.
Area, Ladder Symmetry, Degeneracy and Fluctuations of a Horizon. \emph{arXiv},  arXiv:0711.1879 [hep-th].



%

\bibitem{frittrov}
Frittelli, S.; Lehner, L.; Rovelli, C.
The Complete spectrum of the area from recoupling theory in loop quantum gravity.
{\it Class. Quantum Gravity}  \bfseries1996\mdseries,  {\it 13}, 2921--2932.  
Ashtekar,  A.; Lewandowski, J. 1996 Quantum Theory of Geometry I: Area Operator. \emph{Class. Quantum Gravity \textbf{1997}, \emph{14}, A55--A82}.

\bibitem{ansari2}
Dasgupta, A.
Semiclassical horizons.
\emph{Can.  J.  Phys.}   \textbf{2008},  {\it  86}, 659--662.
\bibitem{1999}
 Major, S.A. Operators for quantized directions. {\it Class. Quantum Gravity} \bfseries1999\m, {\it2016}, 3859--3877.
UWTHPH-1999-28.
\bibitem{calzetta2}
Calzetta, E.; Hu, B.L.
Decoherence of Correlation Histories.   \emph{arXiv},  arXiv:gr-qc/9302013.



\bibitem{wheeler}
Wheeler, J. \emph{Pregeometry: Motivations and Prospects, in Quantum
Theory and Gravitation};  Marlow, A., Ed.; Academic
Press:   New Orleans, LA, USA, 1980;  pp.~1--11.
\bibitem{lorente}
Lorente, M. \emph{“Quantum Process and the Foundation of Relational Theories
of Spacetime”, in Relativity in General}; Diaz Alonso, J., Lorente, M.,  Eds.; 
 Editions Frontieres: Paris, France, 1994;  pp. 297--302.
\bibitem{Stuckey:2000ps} 
Stuckey, W.M.; Silberstein, M.
Uniform Spaces in the Pregeometric Modeling of Quantum Nonlocality.  \emph{arXiv},  arXiv:gr-qc/0003104.





\bibitem{friedmann}
Antonsen, F. Models of Pregeometry.  In  Proceedings of the 2nd Alexander Friedmann International Seminar on Gravitation and Cosmology,   St. Petersburg, Russia, 12--19 September 1993;  pp. 287--304.


\bibitem{terazawa0}
Akama, K.; Terazawa, H.
Pregeometric Origin of the Big Bang.
\emph{Gen.  Relariv.  Gravit.}  \textbf{1983},  {\it 15}, 201--207.
\bibitem{terazawa}
Terazawa, H.
Pregeometry.
INS-429, C81-10-13-1.  


\bibitem{alvarez}
Alvarez, E.; Cespedes, J.; Verdaguer, E.
Quantum metric spaces as a model for pregeometry.
{\it Phys. Rev.  D }  \textbf{1992}, {\it 45}, 2033--2043.  
Conference Proceeding, Conference: C81-10-13.1 (Moscow Quant.Grav.1981:0047), Oct 1981. 7p.



\bibitem{ngdam}
Ng, Y.J.; van Dam, H.
Limitation to quantum measurements of space-time distances.  
\emph{Ann.  N.  Y.  Acad. Sci.}   \textbf{1995}, {\it  755}, 579--584.



\bibitem{bel73} Belinskii, V.A.; Khalatnikov, I.M. {\it Sov. Phys. JETP} {\bf 1973}, {\it 4}, 591--597.


\bibitem{pontz2009} 
Pontzen,  A.; Challinor, A. Linearization of homogeneous, nearly-isotropic cosmological models. {\it \mbox{Class. Quantum Gravity}}  \textbf{2011}, {\it 28}, 185007, doi:10.1088/0264-9381/28/18/185007.

\bibitem{dor71}Doroshkevich, A.G.; Novikov, I.D. Mixmaster Universes and the Cosmological Problem.
{\it Sov.  Astron.}  \textbf{1971}, {\it 14}, 763.



\bibitem{lars2004} \textls[-20]{Andersson, L.; van Elst, H.; Uggla, C. Gowdy phenomenology in scale-invariant variables.
{\it \mbox{Class. Quantum Gravity}}  \textbf{2004}, {\it 21}, S29--S57.}

\bibitem{ales10}
Alesci, E.;  Cianfrani, F.
A new perspective on cosmology in Loop Quantum Gravity.
{\it Europhys.~Lett.} {\bf 2013}, {\it 104},  doi:10.1209/0295-5075/104/10001.


\bibitem{Ashtekar:2013xka} 
Ashtekar, A.
Loop Quantum Gravity and the Planck Regime of Cosmology. In \emph{Fundamental Theories of Physics};  van Beijeren, H., Blanchard, P., Busch, P., Coecke, B., Dieks, D., Dittrich, B., Dürr, D., Durrer, R., Frigg, R., Fuchs, C., et al., Eds.; Springer: Berlin/Heidelberg, Germany,  2014; Volume 177, pp.  323--347. 



\bibitem{asht2006}
Ashtekar,  A.; Pawlowski,  T.; Singh,  P.
Quantum Nature of the Big Bang: An Analytical and Numerical Investigation.
{\it Phys.  Rev.  D} \textbf{2006}, {\it  73}, doi:10.1103/PhysRevD.73.124038. 


\bibitem{pots2000} 
Wilson-Ewing, E.
Anisotropic loop quantum cosmology with self-dual variables.
{\it Phys. Rev. D} \textbf{2016}, {\it 93}, 083502, doi:10.1103/PhysRevD.93.083502.


\bibitem{Gielen:2014ila}
Gielen, S.
Quantum cosmology of (loop) quantum gravity condensates: An example.
{\it Class. Quantum Gravity}  \textbf{2014}, {\it 31}, 155009, doi:10.1088/0264-9381/31/15/155009.

\bibitem{Hu:2005ub}
Hu, B.L.
Can spacetime be a condensate?
\emph{Int.\ J.\ Theor.\ Phys.}  \textbf{2005}, {\it 44}, 1785--1806.


\bibitem{Gielen:2013kla} 
Gielen, S.; Oriti, D.; Sindoni, L.
Cosmology from Group Field Theory Formalism for Quantum Gravity.
{\it \mbox{Phys. Rev. Lett.}}    \textbf{2013},  {\it 111}, 031301, doi:10.1103/PhysRevLett.111.031301.

\bibitem{Kiefer:2011cc}
Kiefer, C.; Kraemer, M.
Quantum Gravitational Contributions to the CMB Anisotropy Spectrum.
{\it \mbox{Phys. Rev. Lett.}}    \textbf{2012},  {\it 108}, 021301, doi:10.1103/PhysRevLett.108.021301.

\bibitem{Kiefer:2008bs}
Kiefer, C.
Quantum geometrodynamics: whence, whither?
\emph{Gen.\ Relativ. Gravit.}  \textbf{2009},  {\it 41},   877--901.

\bibitem{Mueller:2016aov}
Müller, M.P.; Carrozza, S.; Höhn, P.A.
Is the local linearity of space-time inherited from the linearity of probabilities?
\emph{J.\ Phys.\ A}   \textbf{2017}, {\it 50}, 054003, doi:10.1088/1751-8121/aa523b.

\bibitem{Ashtekar:1994wa}
Ashtekar, A.; Lewandowski, J.
Differential geometry on the space of connections via graphs and projective limits.
\emph{J.\ Geom.\ Phys.}   \textbf{1995},  {\it 17},  191--230.


\bibitem{Calzetta:1993qe}
Calzetta, E.; Hu, B.L.
Noise and fluctuations in semiclassical gravity.
{\it Phys. Rev. D}  \textbf{1994},  {\it 49},  6636, doi:10.1103/PhysRevD.49.6636.


\bibitem{Hu:1994dka} 
Hu, B.L.; Sinha, S.
A Fluctuation - dissipation relation for semiclassical cosmology.
{\it Phys. Rev. D}  \textbf{1995},  {\it 51}, 1587, doi:10.1103/PhysRevD.51.1587.

\bibitem{ashtx}
 Ashtekar, A.; Kaminski, W.; Lewandowski, J. 
Quantum field theory on a cosmological, quantum space-time.
{\it Phys. Rev. D} \textbf{2009}, {\it 79}, 064030, doi:10.1103/PhysRevD.79.064030.

\bibitem{incond1}
Chen, L.; Zhu, J.Y.
Loop quantum cosmology: The horizon problem and the probability of inflation.
{\it \mbox{Phys. Rev. D}}  \textbf{2015}, {\it 92},   084063, doi:10.1103/PhysRevD.92.084063.

\bibitem{incond2}
Zhu,  T.; Wang,  A.;  Cleaver, G.; Kirsten, K.;  Sheng, Q.
Pre-inflationary universe in loop quantum cosmology.  \emph{arXiv}, arXiv:1705.07544 [gr-qc].





\bibitem{borde} 
Borde, A.; Guth, A.H.; Vilenkin, A.
Inflationary space-times are incompletein past directions.
{\it Phys. Rev. Lett.}    \textbf{2003},  {\it 90}, 151301, doi:10.1103/PhysRevLett.90.151301.

\bibitem{Agullo:2012sh} 
Agullo, I.; Ashtekar, A.; Nelson, W.
A Quantum Gravity Extension of the Inflationary Scenario.
{\it \mbox{Phys. Rev. Lett.}}  \textbf{2012}, {\it 109}, 251301, doi:10.1103/PhysRevLett.109.251301.

\bibitem{anto} Antoniadis, I.; Patil, S.P. The Effective Planck Mass and the Scale of Inflation. \emph{arXiv},  arXiv:1410.8845 [hep-th].







\bibitem{gasper} Gasperini, M.
Cosmology and short-distance gravity.
{\it Int. J. Mod. Phys. D}  \textbf{2015}, {\it 24}, 1544003, doi:10.1142/S0218271815440034.

\bibitem{jacob}
Jacobson, T.
Thermodynamics of space-time: The Einstein equation of state.
{\it Phys. Rev. Lett.} \textbf{1995}, {\it 75}, 1260, doi:10.1103/PhysRevLett.75.1260.

\bibitem{moham}
Mohammadi, A.; Ali, A.F.; Golanbari, T.;  Aghamohammadi, A.; Saaidi, K.; Faizal, M.
Inflationary Universe in the Presence of a Minimal Measurable Length. \emph{arXiv}, arXiv:1505.04392 [gr-qc].







\bibitem{Grishchuk:1993ds} 
Grishchuk, L.P.
Quantum effects in cosmology.
{\it Class. Quantum Gravity} \textbf{1993},  {\it 10}, 2449--2478.

\bibitem{Hogan:2003mq} 
Hogan, C.J.
Discrete spectrum of inflationary fluctuations.
{\it Phys. Rev. D}   \textbf{2004}, {\it70}, 083521, doi:10.1103/Phys\linebreak~RevD.70.083521.


\bibitem{Hogan:2005iw} 
Hogan, C.J.
\textls[-25]{Discrete Quantum Spectrum of Observable Correlations From Inflation.  \emph{arXiv}, 	arXiv:astro-ph/0504364.}







\bibitem{Lesgourgues:1996jc} 
Lesgourgues, J.; Polarski, D.; Starobinsky, A.A.
Quantum to classical transition of cosmological perturbations for nonvacuum initial states.
{\it Nucl. Phys. B}   \textbf{1997},  {\it 497}, 479--510.

\bibitem{Polarski:1995jg}
Polarski,  D.;   Starobinsky, A.A.
Semiclassicality and decoherence of cosmological perturbations.
{\it \mbox{Class. Quantum Gravity}}  \textbf{1996},  {\it 13}, 377--392.

\bibitem{Grain:2017dqa} 
Grain, J.; Vennin, V.
Stochastic inflation in phase space: Is slow roll a stochastic attractor?
\emph{JCAP}  \textbf{2017},  {\it 1705},  045, doi:10.1088/1475-7516/2017/05/045.




\bibitem{Enqvist:2012xn}
Enqvist, K.; Lerner, R.N.; Taanila, O.; Tranberg, A.
Spectator field dynamics in de Sitter and curvaton initial conditions.
\emph{JCAP}  \textbf{2012}, {\it 1210}, 052, doi:10.1088/1475-7516/2012/10/052.

\bibitem{Kiefer:2008ku}
Kiefer, C.; Polarski, D.
Why do cosmological perturbations look classical to us?
\emph{Adv.\ Sci.\ Lett.}  \textbf{2009}, {\it 2}, 164--173.

\bibitem{Nelson:2016kjm}
Nelson, E.
Quantum Decoherence During Inflation from Gravitational Nonlinearities.
\emph{JCAP}   \textbf{2016}, {\it 1603}, 022, doi:10.1088/1475-7516/2016/03/022.

\bibitem{Burgess:2006jn}
Burgess, C.P.; Holman, R.; Hoover, D.
Decoherence of inflationary primordial fluctuations. 
{\it Phys. Rev. D}    \textbf{2008}, {\it 77}, 063534.

\bibitem{Calzetta:1995ys}
Calzetta, E.; Hu, B.L.
Quantum fluctuations, decoherence of the mean field, and structure formation in the early universe.
{\it Phys. Rev. D}    \textbf{1995}, {\it 52}, 6770--6788. 

\bibitem{Martin:2015qta} 
Martin, J.;  Vennin, V. 
Quantum Discord of Cosmic Inflation: Can we Show that CMB Anisotropies are of Quantum-Mechanical Origin?
{\it Phys. Rev. D}    \textbf{2016},  {\it 93}, 023505, doi:10.1103/PhysRevD.93.023505.

\bibitem{Markkanen:2016jhg}
Markkanen, T. 
Decoherence Can Relax Cosmic Acceleration.
\emph{JCAP}  \bfseries2016\mdseries, {\it  1611}, 026, doi:10.1088/1475-751\linebreak6/2016/11/026.


\bibitem{Boyanovsky:2015tba}
 Boyanovsky, D.
Effective field theory during inflation: Reduced density matrix and its quantum master equation.
{\it Phys. Rev. D}  \textbf{2015}, {\it 92}, 023527, doi:10.1103/PhysRevD.92.023527.

\bibitem{Gambini:2004bm} 
Gambini, R.; Pullin, J.
Canonical quantum gravity and consistent discretizations.  \emph{Pramana} \textbf{2004}, \emph{63},    755--764.




\bibitem{Chatwin-Davies:2016byj} 
Chatwin-Davies, A.; Kempf, A.; Martin, R.T.W.
Impact of Natural Planck Scale Cutoffs that are Fully Covariant on Inflation.  \emph{	Phys. Rev. Lett.} \textbf{2017}, \emph{119},  doi:10.1103/PhysRevLett.119.031301.



\bibitem{hogan}
Hogan, C.J.   
Quantum Gravitational Uncertainty of Transverse Position. \emph{arXiv},  arXiv:astro-ph/0703775.






\bibitem{macr2017} Jacobson, T.  
Entanglement Equilibrium and the Einstein Equation.  \emph{Phys. Rev. Lett.}  \textbf{2016}, \emph{116},  201101, doi:10.1103/PhysRevLett.116.201101. 
Hogan,~C.J. Measurement of Quantum Fluctuations in Geometry. {\it Phys.~Rev.~D}  \textbf{2008}, {\it 77}, 104031, doi:10.1103/PhysRevD.77.104031. 




\bibitem{hogan4}
Hogan, C.J.
Quantum Indeterminacy of Emergent Spacetime.  \emph{arXiv}, arXiv:0710.4153 [gr-qc].





\bibitem{puzio} 
Puzio, R. On the square root of the Laplace-Beltrami operator as a Hamiltonian. {\it Class. Quantum Gravity}   \textbf{1994}, {\it 11}, 609, doi:10.1088/0264-9381/11/3/013.

\bibitem{hogan3}
Hogan, C.J.
Quantum Gravitational Uncertainty of Transverse Position. \emph{arXiv}, arXiv:astro-ph/0703775.







\bibitem{Major:2011ry} 
Major, S.A.
Quantum Geometry Phenomenology: Angle and Semiclassical States.
{\it J. Phys. Conf. Ser.} \p2012\m, {\it 360}, 012061, 
doi:10.1088/1742-6596/360/1/012061.

\bibitem{hogan2}
Hogan, C.J.
Spacetime Indeterminacy and Holographic Noise.  \emph{arXiv}, arXiv:0706.1999 [gr-qc].





\bibitem{cardella}
Cardella, M.A.;   Zanon, D.
Noncommutative deformation of four dimensional Einstein gravity. 
\emph{\mbox{Class. Quantum Gravity}} {\bf 2003}, {\it 20},   L95--L104.
Available online: [hep-th/0212071].
		

	
	
	
	
	
\bibitem{emergent}
Verlinde, E.P.
On the Origin of Gravity and the Laws of Newton.
\emph{JHEP}   \textbf{2011},  {\it 1104}, 029, doi:10.1007/JHEP04(2011)029.
Available online: [arxiv:1001.0785 [hep-th]].

\bibitem{hogan5}
Chou, A.;    Glass, H.;    Gustafson, H.R.;   Hogan, C.J.;  Kamai, B.L.;   Kwon, O.;   Lanza, R.;  McCuller, L.;   Meyer,~S.S.;   Richardson, J.W.; et al.
Interferometric Constraints on Quantum Geometrical Shear Noise Correlations.  \emph{arXiv}, arXiv:1703.08503 [gr-qc].







\bibitem{hogan6}
Hogan, C.J.; Kwon, O.
Statistical Measures of Planck Scale Signal Correlations in Interferometers.
{\it \mbox{Class. Quantum Gravity}}   \textbf{2017},  {\it 34},  075006, doi:10.1088/1361-6382/aa601e.

\bibitem{hogan9}
Hogan, C.J.
Interferometers as Probes of Planckian Quantum Geometry.
{\it Phys. Rev. D}    \textbf{2012},  {\it 85}, 064007, doi:10.1103/PhysRevD.85.064007

\bibitem{hogan10}
\textls[-20]{Kwon, O.; Hogan, C.J.
Interferometric Tests of Planckian Quantum Geometry Models.
{\it \mbox{Class. Quantum Gravity}} } \textbf{2016},  {\it 33},  105004, doi:10.1088/0264-9381/33/10/105004.

\bibitem{alwis}
De Alwis, S.P.
Cosmological Fluctuations: Comparing Quantum and Classical Statistical and Stringy Effects.  \emph{arXiv}, arXiv:1504.05211 [hep-th].








\bibitem{bunch}
Bunch, T.S.; Davies, P.C.W.  Quantum Field Theory in De Sitter Space: Renormalization by Point Splitting. 
{\it Proc. R. Soc. Lond. A}  \textbf{1978}, {\it 360}, 117--134. 

\bibitem{wein} Weinberg, S. Quantum contributions to cosmological correlations.  {\it Phys. Rev. D}  \textbf{2005}, {\it 72}, 043514, doi:10.1103/PhysRevD.72.043514.




\bibitem{Ng:1994zk}
Ng, J.N.; van Dam, H.
Limitation to quantum measurements of space-time distances.
\emph{Ann. N.\ Y.\ Acad.\ Sci.}   \textbf{1995}, {\it 755}, 579--584.


\bibitem{Ng:1993jb} 
Ng, Y.N.; Van Dam, H.
Limit to space-time measurement.
{\it Mod. Phys. Lett.  A}   \textbf{1994},  {\it 9}, 335, doi:10.1142/S0217732394000356.

\bibitem{Salecker:1957be} 
Salecker, H.; Wigner, E.P.
Quantum limitations of the measurement of space-time distances.
{\it Phys. Rev.}  \textbf{1958},  {\it 109}, 571--577.

\bibitem{AmelinoCamelia:1994vs} 
Amelino-Camelia, G.
Limits on the measurability of space-time distances in the semiclassical approximation of quantum gravity.
{\it Mod. Phys. Lett. A}   \textbf{1994},  {\it 9}, 3415--3422.

\bibitem{amel}
Amelino-Camelia, G.
A Phenomenological description of quantum gravity induced space-time noise.
\emph{Nature}  \textbf{2001},  {\it 410}, 1065--1067.

\bibitem{amel2}
Amelino-Camelia, G.
Phenomenological Description of Space-Time Foam.   \emph{arXiv}, arXiv:gr-qc/0104005.







\bibitem{calmet}
Calmet, X.
On the Precision of a Length Measurement.
\emph{Eur. Phys. J. C}   \textbf{2008},  {\it 54}, 501--505.
[Subnucl. Ser.  \textbf{2008}, {\it 44}, 625].   



\bibitem{Requardt:2008yy}
Requardt, M.
About the Minimal Resolution of Space-Time Grains in Experimental Quantum Gravity.  \emph{arXiv}, arXiv:0807.3619 [gr-qc].







\bibitem{Adler:1999if} 
Adler, R.J.; Nemenman, I.M.; Overduin, J.M.; Santiago, D.I.
On the detectability of quantum space-time foam with gravitational wave interferometers.
{\it Phys. Lett. B}   \textbf{2000},  {\it 477}, 424--428.

\bibitem{Baez:2002ra} 
Baez, J.C.; Olson, S.J.
Uncertainty in measurements of distance.
{\it Class. Quantum Gravity}   \textbf{2002},  {\it 19}, L121--L126.

\bibitem{Requardt:2005yw}
Requardt, M.
Planck fluctuations, measurement uncertainties and the holographic principle.
{\it Mod. Phys. Lett.~A}    \textbf{2007}, {\it 22}, 791--806.


\bibitem{Cahill:2002gs} 
Cahill, R.T.  
Process Physics: From Quantum Foam to General Relativity. \emph{arXiv}, arXiv:gr-qc/0203015.






\bibitem{gran2}
Burgio, G.; De Pietri, R.; Morales-Tecotl, H.; Urrutia, L.F.; Vergara, J.D.
The Basis of the physical Hilbert space of lattice gauge theories.
{\it Nucl. Phys. B}   \textbf{2000},  {\it 566}, 547--561. 


\bibitem{gibbs1}
Gibbs, P.
The Small Scale Structure of Space-Time: A Bibliographical Review. \emph{arXiv}, arXiv:hep-th/9506171.





\bibitem{gibbs2}
Zhang, A.W.
Mathematical Structure of Discrete Space-time.  \emph{arXiv}, arXiv:0910.1417 [quant-ph].








\bibitem{pregeo2}
Cahill, R.T.;  Klinger, C.M. 
Pregeometric modeling of the space-time phenomenology.
{\it Phys. Lett. A}   \textbf{1996},  {\it 223}, 313--319.

\bibitem{Henson:2006kf} 
Henson, J.
The Causal set approach to quantum gravity.
In  \emph{Approaches to Quantum Gravity}; Oriti, D., Ed.;  Cambridge University Press: Cambridge, UK, 2009; pp. 393--413.




\bibitem{Bombelli:1989mu} 
Bombelli, L.; Meyer, D.A.
The Origin of Lorentzian Geometry.
{\it Phys. Lett. A}   \textbf{1989},  {\it 141}, 226--228.



\bibitem{Pramanik:2014mda}
  Pramanik, S. Implication of the geodesic equation in the generalized uncertainty principle framework.
  {\it Phys. Rev. D} \textbf{2014} {\it 90},  024023

\bibitem{Sorkin:2011sp}
  Sorkin, R.D. Toward a 'fundamental theorem of quantal measure theory'.
  Available online: arXiv:1104.0997 [hep-th].
	
	
	\bibitem{Kempf:2012sg}
  Kempf, A., Chatwin-Davies, A., Martin, R.T.W. A fully covariant information-theoretic ultraviolet cutoff? for scalar ?fields in expanding FRW spacetimes.
  {\it J. Math. Phys.}, {\textbf{2013}, {\it 54}, 022301
  Available online: [arXiv:1210.0750 [gr-qc]].
	
	\bibitem{Mewes:2013cda}
  Mewes, M.
  Higher-order Lorentz violation.
  Available online: arXiv:1307.7969 [hep-ph].
	
	
	\bibitem{Shankaranarayanan:2004dk}
  Shankaranarayanan, S., Sriramkumar, L.
  Planck scale effects and the suppression of power on the large scales in the primordial spectrum.
  Available online: hep-th/0410072.
	
	\bibitem{Kaufman:2014rpa}
  Kaufman, J.P., Keating, B.G., Johnson, B.R. Precision Tests of Parity Violation over Cosmological Distances.
  {\it Mon. Not. Roy. Astron. Soc.}, {\textbf 2016}, {\it 455},  1981.
  Available online: [arXiv:1409.8242 [astro-ph.CO]].
	
	\bibitem{Halpern:1999ng}
  Halpern, M., Scott, D.
  Future cosmic microwave background experiments.
  Available online: astro-ph/9904188.
	
	\bibitem{HernandezMonteagudo:2002df}
  Hernandez-Monteagudo, C., Kashlinsky, A., Atrio-Barandela, F.
  Using peak distribution of the cosmic microwave background for WMAP and Planck data analysis: Formalism and simulations.
  {\it Astron. Astrophys.} , {\textbf 2004}, {\it 413},  833
  Available online: [astro-ph/0204458].



\bibitem{gran1}
Urrutia, L.F.
Corrections to flat-space particle dynamics arising from space granularity.
\emph{Lect. Notes Phys.} \textbf{2006},  {\it 702}, 299--345.


\bibitem{gran5}
Akama, K.; Oda, I.
Topological pregauge pregeometry.
\emph{Phys. Lett.  B}   \textbf{1991},  {\it 259}, 431--435.

\bibitem{gran3}
Aguilar, P.; Sudarsky, D.; Bonder, Y.
Experimental search for a Lorentz invariant spacetime granularity: Possibilities and bounds.
{\it Phys. Rev. D} \textbf{2013},  {\it 87},   064007, doi:10.1103/PhysRevD.87.064007.

\bibitem{gran4}
Bonder, Y.
Lorentz violation in a uniform newtonian gravitational field.
{\it Phys. Rev. D}     \textbf{2013},  {\it 88}, 105011, doi:10.1103/PhysRevD.88.105011.


\bibitem{pregeo1}
Meschini, D.; Lehto, M.; Piilonen, J.
Geometry, pregeometry and beyond.
\emph{Stud.\ Hist.\ Philos.\ Sci.\ B }  \textbf{2005},  {\it 36}, 435--464.


\bibitem{sark}
Sarkar, S.
Possible astrophysical probes of quantum gravity.
{\it Mod. Phys. Lett. A}    \textbf{2002},  {\it 17}, 1025--1036.


\bibitem{sark2}
Jacobson, T.; Liberati, S.; Mattingly, D.
Threshold effects and Planck scale Lorentz violation: Combined constraints from high-energy astrophysics.
{\it Phys. Rev. D}   \textbf{2003}, {\it 67}, 124011, doi:10.1103/PhysRevD.67.124011.

\bibitem{sark3}
{Mattingly, D. 
Have we tested Lorentz invariance enough?  \emph{arXiv}, arXiv:0802.1561 [gr-qc].}








\bibitem{fant2014a}
Belenchia, A.; Benincasa, D.M.; Martin-Martinez, E.; Saravani, M. Low energy signatures of nonlocal field theories. {\it Phys. Rev. D}  \textbf{2016}, {\it 94} 061902, doi:10.1103/PhysRevD.94.061902.


\bibitem{Girelli:2012ju}
Girelli, F.; Hinterleitner, F.; Major, S.
Loop Quantum Gravity Phenomenology: Linking Loops to Observational Physics.
\emph{SIGMA}   \textbf{2012},  {\it 8}, 098, doi:10.3842/SIGMA.2012.098.

\bibitem{Agullo:2013ai} 
Agullo, I.; Ashtekar, A.; Nelson, W.
The pre-inflationary dynamics of loop quantum cosmology: Confronting quantum gravity with observations.
{\it Class. Quantum Gravity}  \textbf{2013},  {\it 30}, 085014, doi:10.1088/026\linebreak4-9381/30/8/085014.

\bibitem{Wilson-Ewing:2016yan}
Wilson-Ewing, E.
Testing loop quantum cosmology.
\emph{Comptes Rendus Phys.}  \textbf{2017}, {\it 18},  207--225.

\bibitem{grish}
Grishchuk, L.P.
Signatures of quantum gravity in the large scale universe.
In {\it  Phase Transitions in Cosmology}, \emph{Proceedings, 4th Cosmology Colloquium, Euroconference, Paris, France,  4--9 June 1997};     pp. 176--195.  


\bibitem{Sung:2010ek} 
Sung, R., Coles, P. Temperature and Polarization Patterns in Anisotropic Cosmologies. \emph{JCAP}  \textbf{2011}, \emph{1106} , 036, doi:10.1088/1475-7516/2011/06/036.


\bibitem{chall} Pontzen, A.; Challinor, A.
Bianchi Model CMB Polarization and its Implications for CMB Anomalies.
{\it Mon. Not. R. Astron. Soc.}  \textbf{2007}, {\it 380}, 1387--1398.


\bibitem{king2006}
King, E.; Coles, P.  
Dynamics of a magnetised Bianchi I universe with vacuum energy. {\it Class. Quantum Gravity}  \textbf{2007}, {\it 24}, 2061--2072.


\bibitem{rockhee2010} 
Sung, R.; Short, J.; Coles, P. Statistical Characterization of Temperature Patterns in Anisotropic Cosmologies. {\it Mon. Not. R. Astron. Soc.}  \textbf{2011}, {\it 412} 492,  doi:10.1111/j.1365-2966.2010.17922.x.




\bibitem{Rockhee2009}
Sung, R.; Coles, P.
Polarized Spots in Anisotropic Open Universes.
{\it Class. Quantum Gravity} \textbf{2009}, \emph{26},   172001, doi:10.1088/0264-9381/26/17/172001.



	\bibitem{Cho:2003af}
  Cho. J.Y., Lazarian, A. Generation of compressible modes in MHD turbulence.
  {\it Theor. Comput. Fluid Dyn.}, {\textbf 2005},  {\it 19},  127
  doi:10.1007/s00162-004-0157-x
 Available online:  [astro-ph/0301462].
	
	
	\bibitem{ichik14}
	
Ichiki, K.
CMB foreground: A concise review. {\it Progr. of Theor. and Exp. Phys.},
{\textbf 2014}, {\it 6}, 06B109,


		
\bibitem{Adam:2015tpy}
  Adam, R. {\it et al.} [Planck Collaboration]. Planck 2015 results. IX. Diffuse component separation: CMB maps.
  {\it Astron. Astrophys.}, {\textbf 2016}, {\it A9}, 594.
  doi:10.1051/0004-6361/201525936
  Available online: [arXiv:1502.05956 [astro-ph.CO]].
	
\bibitem{Adam:2015wua}
  Adam, R., {\it et al.} [Planck Collaboration].
  Planck 2015 results. X. Diffuse component separation: Foreground maps.
  {\it Astron. Astrophys.}, {\textbf 2016}, {\it 594},  A10, 63 pp.,
  Available online: [arXiv:1502.01588 [astro-ph.CO]].
	

	
\bibitem{Ade:2015qkp}
  Ade, P.A.R. {\it et al.} [Planck Collaboration]. Planck 2015 results. XXV. Diffuse low-frequency Galactic foregrounds.
  {\it Astron. Astrophys.}  {\bf 594} (2016) A25
  Available online: [arXiv:1506.06660 [astro-ph.GA]].
	
	

\bibitem{deOliveiraCosta:2002ng}
  de Oliveira-Costa, A., Tegmark, M., Zaldarriaga, M., Barkats, D., Gundersen, J.O., Hedman, M.M., Staggs, S.T., Winstein, B.
  First attempt at measuring the CMB cross-polarization.
  {\it Phys. Rev. D}, {\textbf 2003}, {\it 67},  023003.
  Available online: [astro-ph/0204021].

	
\bibitem{Choi:2015xha}
  Choi, S.K., Page, L.A.,
  Polarized galactic synchrotron and dust emission and their correlation.
  {\it JCAP} {\textbf 2015}, {\it 1512},  020
  Available online: [arXiv:1509.05934 [astro-ph.CO]].
	
\bibitem{Cho:2010kw}
  Cho, J., Lazarian, A. Galactic foregrounds: Spatial fluctuations and a procedure of removal.
  {\it Astrophys. J.}, {\textbf 2010}, {\it 720}, 1181
  doi:10.1088/0004-637X/720/2/1181
  Available online: [arXiv:1007.3740 [astro-ph.CO]].
	
\bibitem{Yamauchi:2016ypt}
	  Yamauchi, D., {\it et al.}.5 [SKA-Japan Consortium Cosmology Science Working Group],
  Cosmology with the Square Kilometre Array by SKA-Japan.
  {\it PoS DSU}, {\textbf 2016}, {\it 2015}, 004.
   [Publ.\ Astron.\ Soc.\ Jap.\  {\bf 68} (2016) no.6,  R2]
  Available online: [arXiv:1603.01959 [astro-ph.CO]].
\bibitem{Ruud:2015pvs}
  Ruud. T.M., {\it et al.}
  The Q/U Imaging ExperimenT: Polarization Measurements of the Galactic Plane at 43 and 95 GHz.
  {\it Astrophys. J.}, {\textbf 2015},  {\it 811}, 89.
  Available online: [arXiv:1508.02778 [astro-ph.CO]].
\bibitem{Namikawa:2015tba}
  Namikawa, T., Nagata, R. Non-Gaussian Structure of B-mode Polarization after Delensing.
  {\it JCAP}, {\textbf 2015}, {\it 1510}, 004.
  doi:10.1088/1475-7516/2015/10/004
  Available online: [arXiv:1506.09209 [astro-ph.CO]].
	
\bibitem{planck2013}
Ade, P.A.R.,   Aghanim, N.; Armitage-Caplan, C.,  Arnaud, M.,  Ashdown, M.,  Atrio-Barandela, F., Aumont, J.,  Baccigalupi, C.,  Banday, A.J.,  Barreiro, R.B., et al.   [Planck Collaboration]. 
\emph{Astron.  Astrophys.} \textbf{2014},  {\it 571}, A26, doi:10.1051/0004-6361/201321546. 


\bibitem{Purkayastha:2017sjj}
  Purkayastha, U., Saha, R., Separating CMB Stokes Q and U polarization signals from Non-Gaussian Emissions.
  Available online: arXiv:1707.02008 [astro-ph.CO].
\bibitem{diSeregoAlighieri:2016lbr}
  di Serego Alighieri, S. The conventions for the polarization angle.
  {\it Exper. Astron.}, {\textbf 2017}, {\it 43},  19.
  Available online: [arXiv:1612.03045 [astro-ph.IM]].
\bibitem{planck2015} 
Ade,  P.A.R.; Aghanim, N.; Armitage-Caplan, C.;  Arnaud, M.;  Ashdown, M.;   Atrio-Barandela, F.; Aumont,  J.;  Baccigalupi, C.; Banday,  A.J.;   Barreiro, R.B.; et al.
Planck Collaboration, Planck 2015 results. XVI. Isotropy and statistics of the CMB. {\it Astron.   Asrtophs}  \textbf{2016}, {\it 594},  A16, doi:10.1051/0004-6361/201321546.

\bibitem{fant2014} Fantaye, Y. Test of cosmic isotropy in the Planck era.
In {Proceedings of the Invited Talk at the Cosmology Session of `$49$th Rencontres de Moriond $2014$'}, La Thuile,  Italy,   22--29 March 2014.
	\bibitem{Zhai:2017ibd}
  Z.~Zhai and M.~Blanton,
  Available online: arXiv:1707.06555 [astro-ph.CO].
\bibitem{Gorski:2004by}
  Gorski, K.M., Hivon, E., Banday, A.J., Wandelt, B.D., Hansen, F.K., Reinecke, M., Bartelman, M. HEALPix - A Framework for high resolution discretization, and fast analysis of data distributed on the sphere.
  {\it Astrophys. J.}, {\textbf 2005}, {\it 622},  759
  Available online: [astro-ph/0409513].
\bibitem{Bucher:2015eia}
  Bucher, M., Physics of the cosmic microwave background anisotropy.
  {\it Int. J. Mod. Phys. D}, {\textbf 2015}; {\it 24},  1530004.
  Available online: [arXiv:1501.04288 [astro-ph.CO]].

\bibitem{Anber:2011ut}
Anber,    M.M.; Donoghue, J.F.
On the running of the gravitational constant.
{\it{Phys. Rev. D}} {\bf 2012}, {\it 85}, 104016, doi:10.1103/PhysRevD.85.104016.
Available online: [arXiv:1111.2875 [hep-th]].


\bibitem{BjerrumBohr:2002kt} 
Bjerrum-Bohr, N.E.J.; Donoghue, J.F.; Holstein, B.R. Quantum gravitational corrections to the nonrelativistic scattering potential of two masses.
{\it{Phys. Rev. D}} {\bf 2003} {\it 67}, 084033,  doi:10.1103/PhysRevD.67.084033.

\bibitem{Khriplovich:2002bt}
Khriplovich, I.B.; Kirilin, G.G.
Quantum power correction to the Newton law.
{\it{J. Exp. Theor. Phys.}} {\bf 2002}, {\it 95}, 981--986.

\bibitem{stelle1977}
Stelle, K.S. Renormalization of higher-derivative quantum gravity.
{\it{ Phys. Rev. D}}
{\bf{1977}}, {\it{16}}, 953--969


\bibitem{Conroy:2017nkc} 
Conroy, A.; Edholm, J.
Newtonian Potential and Geodesic Completeness in Infinite Derivative Gravity.
{\it \mbox{Phys. Rev. D}} {\bf 2017},  {\it  96}, 044012, doi:10.1103/PhysRevD.96.044012.






\bibitem{Prugovecki:1989kz}
 Prugovecki, E.
Generally Covariant Geometrostochastic Quantum Gravity.
{\it Nuovo Cimento   A} \textbf{1989}, {\it 102},  881--923.

\bibitem{dre1}  Drechsler, W.;  Prugovečki, E. 
Geometrostochastic quantization of a theory for extended elementary objects.  \emph{Found. Phys.}   \textbf{1991}, \emph{21}, 513--546. MPI-PAE-PTH-18-90.

\bibitem{dre2}   Drechsler, W.
Quantized fiber dynamics for extended elementary objects involving gravitation.  \emph{Found. Phys.} \textbf{1992},  \emph{22}, 1041--1077.  MPI-PH-91-33.

\bibitem{anandan}
Anandan, J.S. 
Classical and Quantum Physical Geometry. \emph{arXiv}, arXiv:gr-qc/9712015.





\bibitem{gac2015}  \textls[-15]{Amelino-Camelia,  G.; da Silva, M.M.;  Ronco, M.; Cesarini, L.; Lecian, O.M. 
Spacetime-noncommutativity regime of Loop Quantum Gravity. \emph{Phys. Rev. D} \textbf{2017}, \emph{95}, 024028, doi:10.1103/PhysRevD.95.024028.}



\bibitem{carlip0909} Carlip, S.  Spontaneous Dimensional Reduction in Short-Distance Quantum Gravity?    \emph{arXiv}, arxiv:0909.3329.















\bibitem{bojo2013} Bojowald, M.; Paily, G.M. Deformed General Relativity. {\it Phys. Rev. D}  \textbf{2013}, {\it  87}, 044044, doi:10.1103/PhysRev\linebreak~D.87.044044.

\bibitem{kova2001} Kovacevic, D.; Meljanac, S.; Pachol, A.;   Strajn, R. Generalized Poincare algebras, Hopf algebras and kappa-Minkowski spacetime.  \emph{Phys. Lett. B}   \textbf{2012}, \emph{711},  122--127. 

\bibitem{miel2014}  Mielczarek, J.  Loop-deformed Poincare algebra. {\it EPL} \textbf{2014}, {\it 108}, 40003, doi:10.1209/0295-5075/108/40003. 





\bibitem{lars2015} Larson, D.; Weiland, J.L.; Hinshaw, G.; Bennett, C.L. Comparing Planck and WMAP: Maps, Spectra, and Parameters.
{\it Astrophys. J.}   \textbf{2015}, {\it 801}, 9,  doi:10.1088/0004-637X/801/1/9.
Available online: [arxiv:1409.7718].}
\end{thebibliography}
\renewcommand\bibname{References}

\end{document}